\documentclass[12pt]{article}
\newcommand{\offlatexfilter}{\index{offlatexfilter}}
\offlatexfilter
\newcommand{\onlatexfilter}{\index{onlatexfilter}}
\offlatexfilter
\newcommand{\ignore}[1]{\index{ignore}}
\usepackage[width=6.5in,letterpaper]{geometry}                
\usepackage{ccaption} 

\usepackage{floatpag} 
\usepackage{graphicx}
\usepackage{amsmath}
\usepackage{amsfonts}
\usepackage{amssymb}
\usepackage{epstopdf}
\usepackage{float}

\usepackage[apaciteclassic,nodoi]{apacite}

\graphicspath{
  {/Users/yashar/Dropbox/Documents/Tex/Figures/}
  {figs/}
  }

\usepackage{bm}

\def\vectwo#1#2{\left(\begin{array}{c}  #1 \\ #2 \end{array}\right)}

\def\mat#1#2#3#4{\left(\begin{array}{cc} #1 & #2 \\ #3 & #4 \end{array}\right)}

\def\msub#1{{\mathop{\rm #1}}}

\def\r{{\mathbf r}}
\def\re{{\mathbf r}_E}
\def\ri{{\mathbf r}_I}
\def\drdt{\frac{d\r}{dt}}
\def\y{{\mathbf y}}

\def\dydt{\frac{d\y}{dt}}
\def\h{{\mathbf h}}
\def\he{{\mathbf h}_E}
\def\hi{{\mathbf h}_I}
\def\g{{\mathbf g}}

\def\f{{\mathbf f}}

\def\vv{{\mathbf v}}
\def\W{{\mathbf W}}
\def\Wee{{\mathbf W}_{EE}}
\def\Wei{{\mathbf W}_{EI}}
\def\Wie{{\mathbf W}_{IE}}
\def\Wii{{\mathbf W}_{II}}
\def\Wxy{{\mathbf W}_{XY}}
\def\J{{\mathbf J}}
\def\A{{\mathbf A}}
\def\B{{\mathbf B}}

\def\bPhi{{\bm \Phi}}
\def\bPhir{{\bm \Phi_{\r}}}
\def\Jee{J_{EE}}
\def\Jei{J_{EI}}
\def\Jie{J_{IE}}
\def\Jii{J_{II}}

\def\T{{\mathbf T}}
\def\M{{\mathbf M}}
\def\1{{\mathbf 1}}
\def\eg{{\it e.g.}}
\def\ie{{\it i.e.}}

\def\Det{{\text{Det}}}
\def\Tr{{\text{Tr}\,}}
\def\deg{{\mathrm{o}}}

\newcommand{\be}{\begin{equation}}
\newcommand{\ee}{\end{equation}}
\newcommand{\bea}{\begin{eqnarray}}
\newcommand{\eea}{\end{eqnarray}}
\newcommand{\bs}{\begin{split}}
\newcommand{\bes}{\begin{equation}\begin{split}}
\newcommand{\ees}{\end{split} \end{equation}}
\newcommand{\es}{\end{split}}

\newcommand{\lpr}{\left(}
\newcommand{\rpr}{\right)}
\newcommand{\lbr}{\left[}
\newcommand{\rbr}{\right]}

\newcommand{\areq}{&=&}

\newcommand{\req}[1]{Eq.~(\ref{#1})}
\newcommand{\reqs}[1]{Eqs.~(\ref{#1})}
\newcommand{\myref}[1]{(\ref{#1})}

\newcommand{\vc}[1]{{\mathbf{#1}}}
\newcommand{\vct}[1]{{\boldsymbol{#1}}}

\newcommand{\one}{\mathbf{1}}

\newcommand{\vr}{\vct{r}}
\newcommand{\vy}{\vc{y}}

\newcommand{\vg}{\vct{g}}

\newcommand{\tJ}{{\mathbf{K}}}
\newcommand{\calJ}{\mathcal{J}}

\newcommand{\bPhiy}{\bPhi}

\newcommand{\xx}{x}

\title{Analysis of the stabilized supralinear network$^{\ddagger}$}
\author{Yashar Ahmadian$^{1,4,*}$, Daniel  B.\ Rubin$^{1,2, *}$ \& Kenneth D.\ Miller$^{1-4, \dagger}$}

\begin{document}
\offlatexfilter
\maketitle


{ \setlength{\parindent}{0in}%
  \slshape\sloppy%
  $^1$Center for Theoretical Neuroscience, $^1$Dept. of Neuroscience, $^2$Doctoral Program in
  Neurobiology and Behavior, $^3$Swartz
  Program in Theoretical Neuroscience, and $^4$Kavli Institute for
  Brain Science, College of Physicians and Surgeons, Columbia
  University, NY, NY 10032.\\
$^* $: These authors contributed equally to this work.\\
$^\dagger$: To whom correspondence should be addressed: ken@neurotheory.columbia.edu.\\
 $^ \ddagger $: This preprint has been published as a peer-reviewed article:}\\   Ahmadian, Y., Rubin, D.~B., \& Miller, K.~D. (2013). {Analysis of the stabilized supralinear network.} \emph{Neural Computation, 25}, 1994--2037.
 {\footnotesize (This preprint differs from the published article in that here Figs.~1 and 4  are in color, and we have added ``$k=0.04$" to the caption of Fig.~2.)}

\par\vspace{12pt}

\onlatexfilter
\begin{center}
{\large \bf Abstract}
\end{center}

\begin{quote}

We study a rate-model neural network composed of excitatory and
inhibitory neurons in which neuronal input-output functions are power
laws with a power greater than 1, as observed in primary visual
cortex.  This supralinear input-output function leads to supralinear
summation of network responses to multiple inputs for weak inputs.  We
show that for stronger inputs, which would drive the excitatory
subnetwork to instability, the network will dynamically stabilize
provided feedback inhibition is sufficiently strong.  
For a wide range of network and stimulus parameters,  
this dynamic
stabilization yields a transition from supralinear 
to sublinear summation of network responses to multiple inputs. 
We compare this to
the dynamic stabilization in the ``balanced network'', which yields
only linear behavior.  We more exhaustively analyze the 2-dimensional
case of 1 excitatory and 1 inhibitory population. We show that in this
case dynamic stabilization will occur whenever the determinant of the
weight matrix is positive and the inhibitory time constant is
sufficiently small, and analyze the conditions for
``supersaturation'', or decrease of firing rates with increasing
stimulus contrast (which represents increasing input firing rates). In
work to be presented elsewhere, we have found that this transition from
supralinear to sublinear summation can explain a wide variety of nonlinearities in cerebral cortical processing.  

\end{quote}

\vspace{2\baselineskip}
\noindent {\bf Acknowledgements:} 
D.B.R.\ is supported by NIH training grant T32-GM007367 to the
M.D./Ph.D. training program at Columbia University.  Y.A.\ is
supported by a postdoctoral fellowship from the Kavli Institute for
Brain Science at Columbia University.  K.D.M. is supported by R01
EY11001 from the NEI of the NIH and by the Gatsby Charitable
Foundation through the Gatsby Initiative in Brain Circuitry at
Columbia University.

\newpage

\tableofcontents

\newpage

\section{Introduction}

In work to be presented elsewhere (presented as abstracts in \citeA{Miller_Rubin10,Rubin_Miller10,Miller_Rubin11,Rubin_Miller11}),
we have found that a large set of response properties of cells in primary visual
cortex (V1) and other sensory cortical areas can be understood from a
very simple circuit motif.  The response properties have in common a
change in integration with increasing input strength, so that
responses to weak inputs sum supralinearly while those to stronger
inputs generically sum sublinearly.  In this paper, we mathematically
analyze the model's behavior.

One set of properties the model can address involves contextual modulation or ``surround
suppression''.  A visual sensory neuron has a classical receptive
field (CRF), corresponding to the region in which appropriate visual
stimuli will drive the neuron's responses.  The size of the CRF does
not change with input strength \cite{Song_Li08}.  Stimuli outside the
CRF can modulate responses to CRF stimuli, although they cannot drive
responses, and typically are suppressive.  However, the nature of the
surround influence can vary with input strength
\cite{Sengpiel_etal97,Polat_etal98}.  A size tuning curve is obtained
by centering an effective stimulus on the CRF center and studying
response vs.\ stimulus radius.  The summation field size is the
stimulus size evoking peak response.  This summation field size
shrinks with input strength, as represented by stimulus contrast
\cite{Sceniak_etal99,Cavanaugh_etal02a,Anderson_etal01,Shushruth_etal09,Tsui_Pack11}.
This means that regions of the surround are changing from facilitating
to suppressing with increasing input strength.

Another set of properties involves sublinear summation of the responses
to multiple stimuli: the response to two simultaneously
presented stimuli can be closer to the average than the sum of the
responses to the stimuli presented individually. We refer to this
property as ``normalization'', because it is the most prominent of a
set of nonlinear response properties that have been given that name
(reviewed in \citeA{CarandiniHeeger:2012}).  In at least some cases,
this summation becomes supralinear when inputs are weak
\cite{Heuer_Britten02,Ohshiro_etal11}.  If one thinks of surround
suppression as representing the response to simultaneous presentation
of a center stimulus that normally by itself evokes a certain response and a
surround stimulus that normally by itself evokes zero response, then surround
suppression can be thought of as an example of sublinear summation. Similarly,
facilitation by the near surround for weak inputs then represents
supralinear summation.

As we will show elsewhere, these and other response properties can be
understood in some detail from a simple model.  We consider a network
of excitatory ($E$) and inhibitory ($I$) neurons, extended across a 1-D or
2-D space.  The strengths of each type of connection --
$E\Rightarrow E$, $E\Rightarrow I$, $I\Rightarrow E$, $I\Rightarrow I$ --
fall off as functions of cortical distance.  
We are guided by previous results that showed that the inhibition received by cells is decreased when they
are being suppressed by a surround stimulus, relative to their
response to a CRF stimulus alone \cite{Ozeki_etal09}, and
correspondingly that the firing of inhibitory cells, like that
of excitatory cells, is suppressed by surround stimuli
\cite{Song_Li08}.  These results led to the conclusion that the
$E\Rightarrow E$ connections must be sufficiently strong that, when the
network is being driven by a CRF stimulus, they would render the
network unstable in the absence of feedback inhibition
\cite{Tsodyks_etal97,Ozeki_etal09}, a conclusion also supported by other work
\cite{London_etal10}.  We termed such a network an
inhibition-stabilized network or ISN.

We then add to this the fact that individual neurons have a
supralinear, power-law input-output function.  This is based on
intracellular recordings in anesthetized cat primary visual cortex
(V1) that showed that a neuron's instantaneous firing rate is well
described as a power law function of its instantaneous mean voltage
relative to rest (rates and voltages measured in 30 ms bins) with
powers ranging from 2 to 5, and that this holds true over the entire
dynamic range of neuronal response to visual stimuli
\cite{Priebe_etal04,Priebe_Ferster05,Priebe_Ferster06,Finn_etal07}.\footnote{We 
  are assuming that mean voltage is linear in the 
  input. Nonlinearities such as spike-rate-adaptation currents could
  complicate this picture.  We also are ignoring the fact that the
  power increases with contrast, because the noise level decreases
  with contrast \cite{Finn_etal07}, which yields increasing powers
  \cite{Miller_Troyer02,Hansel_vanVreeswijk02}.  However the picture
  we describe in this 
  paper primarily concerns stabilization against the otherwise
  explosive nonlinearity of a supralinear input-output function.
  Thus, the picture should hold so long as the input-output function
  is supralinear over the cell's dynamic range, as expected for
  fluctuation-driven spiking -- the closer the cell is to threshold,
  the greater the increase in spiking driven by a given increment of
  input.}  This power law relationship is predicted on theoretical
grounds when mean input is subthreshold and spiking is driven by input
fluctuations \cite{Miller_Troyer02,Hansel_vanVreeswijk02}, as appears
to be the case in V1 \cite{Anderson_etal00b}.

Here we
mathematically analyze the model. We focus particularly on exposing
the origins of the generic transition in model behavior from supralinear to 
sublinear summation as input strength increases. This typically
corresponds to a transition from a regime in which the excitatory
network is stable by itself to one in which
the excitatory subnetwork by itself is unstable, but is stabilized by
feedback inhibition.  Hence we refer to the network as the stabilized
supralinear network or SSN.  We also conduct a more detailed analysis
of the 2-dimensional case consisting of a single excitatory and a
single inhibitory population.

\section{Setup: Equations for the Supralinear Network}

We take $\r=\vectwo{\re}{\ri}$ to be the $N$-dimensional vector of
neuronal firing rates, ordered so that the top $N_E$ neurons,
represented by $\re$, are all excitatory neurons, and the remaining
$N_I$ neurons, represented by $\ri$ are inhibitory neurons,
$N_E+N_I=N$. (We refer to the units in our model as ``neurons'', but,
as discussed below, the equations represent average firing rates and
so excitatory or inhibitory units may be better understood as
local interconnected groups of excitatory or inhibitory neurons, over
which the average is taken.)  The matrix of connections between the
neurons is $\W=\mat{\Wee}{-\Wei}{\Wie}{-\Wii}$ where $\Wxy$ is the
matrix of connections from neurons of type $Y$ (E or I) to neurons of
type $X$ and has non-negative entries.  The feedforward input to the
neurons in the network is $\h=\vectwo{\he}{\hi}$.

We study the simplest standard firing-rate-model equations (reviewed
in \citeA{Ermentrout_Terman10}, Chapter 11; \citeA{GK02}, Chapter 6; \citeA{DA01} Chapter 7),  in
which a neuron's firing rate approaches a nonlinear
function of its input with first-order dynamics:
\begin{equation}
\tau \T \drdt = -\r + \f(\W\r + \h)
\label{rate}
\end{equation}
Here $\T$ is a diagonal matrix of relative time constants, \ie\ the
time constant of the $i^{th}$ neuron is $\tau T_{ii}$.  $\f$ is a
vector function of a vector argument that acts elementwise on its
argument, $(\f(\vv))_i = f_i(v_i)$, for some scalar functions of a
scalar variable, $f_i$, where $v_i$ is the $i^{th}$ element of $\vv$.
These rate model equations do not capture fast time scales that arise in
spiking networks, and cannot capture synchronization of spikes across
neurons, but tend to be reliable in describing steady states
or slower aspects of dynamics when neurons spike asynchronously.  We
will focus on the steady state and its stability.

We will study the case in which the $f_i$ are identical for all
elements, $f_i\equiv f$, and $f$ is a rectified power law with power
$n>1$:
\begin{equation}
f(x) = k ([x]_+)^n
\label{fequ}
\end{equation}
where $[x]_+=x$, $x>0$; $=0$, otherwise. We will summarize this
by saying
\begin{equation}
\tau \T \drdt = -\r + k(\W\r + \h)^{.n}
\label{requ}
\end{equation}
where $\vv^{.n}$ is the vector with $i^{th}$ element $([v_i]_+)^n$
(the period in the exponent $.n$, based on Matlab notation, is to
indicate that the operation is done element-by-element rather than to
the vector as a whole).  A power-law relation between the
mean input and mean response arises in the case that
spiking is driven by input fluctuations
\cite{Miller_Troyer02,Hansel_vanVreeswijk02}, and similarly it is
observed in V1 as the relation between the trial-averaged mean voltage
and mean response.  Note that the fact, noted in the Introduction, that
the power law holds over the entire dynamic range of visual responses means
that V1 neurons never reach firing rates at which intrinsic saturation of
the input-output function plays a role. Since V1 firing rates in
response to optimal visual stimuli are
typically among the highest seen in cerebral cortex, the same
conclusion is likely to apply to cortex more generally.
For this reason, we model the
input/output function of the neuron simply as a power law without
considering saturating parts of the input/output function.
Model results will only apply when the model
produces firing rates that remain within the non-saturating regime.

We now change variables to dimensionless ones. This allows us to
determine the dimensionless combinations of parameters on which model
behavior depends and in which expansions for small or large values
may be undertaken.  We let $\psi=\|\W\|$ 
where $\|\W\|$ is some matrix norm or other measure of the size of
$\W$, and write $\W=\psi \J$ with $\J$ dimensionless and $\|\J\|=1$.
Similarly we let $c=|\h|$ and write $\h=c\g$ with $\g$ dimensionless
and $|\g|=1$ (again, $|\g|$ indicates some measure of the size of a
vector, \eg\ a vector norm).  Note that $c$ and $\psi\r$
have the same units, so that $\psi\r/c$ is dimensionless,
and that $kc^n$ and $\r$ have the same units, so that
$kc^n/(c/\psi)=kc^{n-1}\psi$ is dimensionless.  We
thus define the dimensionless variable and parameter:
\begin{eqnarray}
\label{scaling}
\y &=& \r\psi/c\\
\alpha &=& k c^{n-1}\psi
\label{dimensionless}
\end{eqnarray}
Then equation \ref{requ} becomes
\begin{equation}
\tau \T \dydt = -\y + \alpha(\J\y + \g)^{.n}
\label{yequ}
\end{equation}
Thus, given $\J,\g,\T$, and $n$, the dynamics depends only on the
single parameter $\alpha$.  

The fact that Eq.~\ref{yequ} has a single $\alpha$ for all neurons is quite
general: if neuron $i$ had parameter $\alpha_i$, this could be
replaced with $\alpha$ by multiplying all weights $J_{ij}$ and inputs
$g_i$ to neuron $i$ by $\left(\frac{\alpha_i}{\alpha}\right)^{1/n}$,
leaving the form of the equation unchanged.  However the fact that the
equation has a single $n$ for all neurons is a real restriction.
Consideration of $n$'s that vary between neurons or between neuron
types remains a question for future study.

Note that we can rewrite Eq.~\ref{requ} as $\tau \T \drdt = -\r +
f(I)$ where $f(x)=x^{.n}$ and the input $I=k^{\frac{1}{n}}\W\r +
k^{\frac{1}{n}}c\g$.  By incorporating the factors of $k$ into the
input $I$, the steady-state rate $f(I)$ becomes of unit magnitude when
the input $I$ is of unit magnitude.  This is a natural scaling for
defining the
effective recurrent weights, $k^{\frac{1}{n}}\W$, and the effective
input strength, $k^{\frac{1}{n}}c$.  Then $\alpha = k c^{n-1}\psi =
\left(k^{\frac{1}{n}}\|\W\|\right)\left(k^{\frac{1}{n}}c\right)^{n-1}$,
that is, $\alpha=(\text{recurrent weight})(\text{feedforward
  strength})^{n-1}$.  Note also that whether the input is dominated by
feedforward input $c\g$ or recurrent input $\W\r$ is determined by the
size and structure of $\y$ for a given $\alpha$ (because
$\W\r+c\g=c(\J\y+\g)$, so that the balance depends only on the relative
sizes of $\J\y$ vs.\ $\g$), and is not impacted at all by the ratio
$c/\psi$, which naively might be thought to determine the
feedforward/recurrent balance.  For a given $\alpha$, this ratio
simply scales $\r$ ($\r=(c/\psi)\y$).

We will focus on the equation for the steady-state:
\begin{equation}
\y = \alpha(\J\y + \g)^{.n}
\label{yss}
\end{equation}
However, in considering stability of the steady state we will need to
use the dynamical equation \ref{yequ}.

\section{Scaling Argument}
\label{sec:scaling}

In this section, we show that the supralinear network generically
makes a transition between responses that scale in two different ways
with $\alpha$. For weak inputs ($\alpha \ll 1$), the net input
(feedforward plus recurrent) to a neuron grows linearly with the
feedforward strength. Because of the supralinear neuronal input/output
function, this yields
supralinear summation of responses to multiple sets of feedforward
inputs. For stronger inputs ($\alpha \gg 1$), the recurrent input
largely cancels the feedforward input, leaving only a net input component that
grows sublinearly with the feedforward input strength. This yields
sublinear response summation for a broad range of parameters.
The transition between these two scaling regimes, which we refer to as
the supralinear and sublinear scaling regimes respectively, occurs for $\alpha$
of order of magnitude 1, for which we 
use the standard notation $\alpha\sim O(1)$.  
We then compare this dynamic input cancellation to 
that in the balanced network model of Van Vreeswijk and Sompolinsky
(\citeyear{vanVreeswijk_Sompolinsky98}).

The transition from supralinear to sublinear scaling is generally
marked by recurrent excitation becoming strong enough to yield
instability and explosive growth of activity on its own, along with
dynamical stabilization of network activity by feedback
inhibition. The effective connection strength between two neurons
tells how much the steady-state postsynaptic rate changes for a given
change in steady-state presynaptic rate.  This is given by the weight
between the neurons times the postsynaptic gain. The gain is the slope
of the input-output function (Eq.~\ref{fequ}), which is monotonically
increasing with the postsynaptic cell's firing rate. That is, given
the steady-state equation $r_i=k(\sum_j W_{ij}r_j+cg_i)^{.n}$, the
effective weight from neuron $j$ to neuron $i$ is
$\frac{dr_i}{dr_j}=nk(\sum_j W_{ij}r_j + cg_i)^{n-1}W_{ij}=
nk^{\frac{1}{n}} r_i^{\frac{n-1}{n}}W_{ij}$, which monotonically
increases with $r_i$. With increasing input, the network responses
typically increase and hence effective connections grow until the
recurrent excitatory-to-excitatory connections become strong enough to
yield instability absent stabilization by feedback inhibition.
The transition from supralinear to sublinear
scaling is generally closely associated with this transition to
potential excitatory instability, as we will see
(Figs.~\ref{parameters}--\ref{parameters-Nd} and Section~\ref{sec:normcriter}).

\subsection{Scaling for small $\alpha$}
\label{subsec:smallalpha}

For $\alpha\ll 1$ we expect the steady state to satisfy $y \approx
\alpha \g^{.n}$, since then the $\J\y$ term is small relative to the
$\g$ term and so adds only a small correction to this solution.  More
generally, we can write a formal expression for the
steady state by starting with Eq.~\ref{yss} and iteratively substituting
$\alpha(\g + \J\y)^{.n}$ for each instance of $\y$, yielding:
\begin{equation}
\y = \alpha(\g + \alpha \J (\g + \alpha \J (\g + \alpha
\J(\ldots)^{.n})^{.n})^{.n})^{.n}
\label{yalpha}
\end{equation}
or, in terms of $\r$,  
\begin{equation}
\r = k(c\g + k\W (c\g + k\W (c\g + k
\W(\ldots)^{.n})^{.n})^{.n})^{.n}
\end{equation}
where the ellipses indicate infinite repetition of the pattern.
Assuming quantities in the parentheses are positive (which they will be for sufficiently small $\alpha$ as we assume all components of $\g$ are positive) so that we can ignore rectification,
Eq.~(\ref{yalpha}) can be converted into an infinite 
series in increasing integer powers of $\alpha$ with dominating
(lowest-order) term 
$\alpha\g^{.n}$.\footnote{This can be done by expanding each power in
  Eq.~\ref{yalpha} as $\alpha(\g+\alpha\J(\ldots)^{.n})^{.n}=
  \alpha(\g^{.n}+n\g^{.(n-1)}.*\alpha\J(\ldots)^{.n}+\frac{n(n-1)}{2}\g^{.(n-2)}
  .* (\alpha\J(\ldots)^{.n}).^2+\ldots$, where $.*$ indicates
  element-by-element multiplication of two vectors to create another
  vector, and then collecting together the terms of each given order
  in $\alpha$.}  The terms multiplying $\alpha^p$ will involve factors
of $\g$ interspersed with $\J$'s, with the sum of the powers on the
$\g$'s equal to $p(n-1)+1$. 
Similarly for $\r$, one obtains a series involving an infinite set of
powers of $c$, with lowest-order term $k(c\g)^{.n}$ and higher-order
terms proportional to $k^p c^{p(n-1)+1}$ and involving a set of $\g$'s
with summed power also equal to $p(n-1)+1$.  If this series converges,
which it will for sufficiently small $\alpha$, it will give a steady
state solution.

Thus, for small $\alpha$, feedforward inputs sum supralinearly to
produce responses (\ie, the response $\r$ depends supralinearly on the
feedforward input $c\g$).  Intuitively: as previously noted, the
effective connection from neuron $j$ to neuron $i$ is
$nk^{\frac{1}{n}} r_i^{\frac{n-1}{n}}W_{ij}$. For small $\alpha$,
$r_i\approx k(c g_i)^n$ so $k^{\frac{1}{n}}
r_i^{\frac{n-1}{n}}W_{ij}\approx \alpha g_i^{n-1} J_{ij}$ is small, that is, effective connections are small.  In this regime the network is
essentially feedforward driven, with small modifications by the weak
effective recurrent connections.  Since individual cells respond
supralinearly to their inputs, the network sums responses
supralinearly.

\subsection{Scaling for large $\alpha$}
\label{sec-hiconscaling}

The scaling $\y \sim \alpha$ (or equivalently $\r \sim c^n$) cannot hold once $\alpha$ is sufficiently large. In particular, 
for sufficiently large $\alpha$, the series in
Eq.~\ref{yalpha} will explode rather than converge. 
Physically, this occurs because inputs are raised to the power $n>1$ to produce responses which in turn feed back in as inputs; once inputs are sufficiently large, this process is explosive, like a nuclear reaction going critical.  We expect this to occur for $\alpha = O(1)$.  Moreover, for $\alpha \gtrsim 1$, the effective recurrent weights typically become strong enough  that the  excitatory subnetwork by itself becomes unstable in the absence of dynamic feedback inhibition. 

For Eq.~\ref{yss} for the steady state to be self-consistent, the dependence of $\alpha(\J\y+\g)^n$ on the leading $\alpha$ must be
cancelled, because otherwise $\y \sim \alpha$, which enters into
$\J\y$ and (dominating over the feedforward input, $\g$, for large enough $\alpha$)   is raised to the $n^{th}$ power to give\footnote{This could
  be avoided if $\J\y=0$ to leading order in $\alpha$, but that
  requires fine tuning, \ie\ it requires $\Det\ \J=0$.} $\y  = \alpha(\J\y+\g)^n \sim
\alpha^{n+1}$, which in turn enters into $\J\y$, and so on -- the infinite
series in powers of $\alpha$ results, which will blow up for
sufficiently large $\alpha$.  Thus to cancel the leading $\alpha$, 
it must be the case that, to leading order in $\alpha$,
$\J\y+\g\sim \alpha^{-\frac{1}{n}}$. This in turn requires that,
to leading order, $\y$ has the same $\alpha$-dependence as $\g$,
$\y\sim \alpha^0$, so that the leading order of $\y$ can cancel the
$\g$ term leaving only terms of order $\alpha^{-\frac{1}{n}}$. 

Thus, if the steady-state \req{yss} is stable, the recurrent input
$\J\y$ must have dynamically adjusted itself to approximately cancel
the feedforward input $\g$, leaving a remainder that becomes smaller
with growing $\alpha$.  We refer to this as dynamic stabilization, for
two reasons. First, mathematically, this dynamic cancellation is
necessary for the existence of a stable steady state for large
$\alpha$. Second, physically, the cancellation typically arises as the
excitatory subnetwork becomes unstable by itself and the network is
dynamically stabilized by feedback inhibition, although as we will
discuss later (Section \ref{sec:normcriter})
there are cases in which the cancellation
arises without excitatory subnetwork instability.
  
Let us define
\begin{equation}
\beta = \alpha^{-\frac{1}{n}}.
\end{equation}
We write $\y=\y_0+\beta \y_1$, with both $\y_0$ and $\y_1$
approximately $O(\beta^0)$ so that $\beta \y_1$ scales approximately
as $\beta$.  It follows from the requirement $\J\y+\g\sim \beta$ that
\begin{eqnarray}
\label{y0Jg}
\y_0 &=& -\J^{-1}\g.
\end{eqnarray}
Substituting $\y=-\J^{-1}\g +\beta \y_1$ in \req {yss}  then yields
\be
 -\J^{-1}\g+\beta \y_1 = (\J\y_1)^{.n}.
\label{y1equ}
\ee
The latter equation shows that $\y_1$ itself has some further dependence 
on $\beta$. We will further discuss below whether the assumption holds
that $\beta \y_1$ scales approximately as $\beta$.

These arguments can be translated in terms of $\r$. Once $c$ is
sufficiently large, self-consistency requires cancellation of the linear
dependence of $\W\r+c\g$ on $c$, because otherwise $\r\sim c^n$, which
enters back into $\W\r$ and is raised to the $n$ to yield $c^{n^2}$
dependence, and so on.  Cancellation requires that, to leading order,
$\r\sim c$, which in turn requires that to leading order $\W\r+c\g
\sim c^{\frac{1}{n}}$, that is, the net input to a cell grows sublinearly
with
the feedforward input strength.  Writing $\r=c \r_0+ c^{\frac{1}{n}}\r_1$, with
$\r_0$ and $\r_1$ both $O(1)$, we
find that $\r_0=\frac{1}{\psi}\y_0=-\W^{-1}\g$ and
$\r_1=\frac{c}{\psi}\frac{\beta}{c^{\frac{1}{n}}}\y_1 =
\frac{1}{\psi\left(k\psi\right)^{\frac{1}{n}}}\y_1$, with $\r_1$
satisfying $-\W^{-1}\g+c^{-\frac{n-1}{n}}\r_1=\left(\W\r_1\right)^n$.

These solutions show that, if the network dynamically stabilizes, the
net input to cells grows sublinearly with the feedforward input strength and
responses are given by a sum of terms that are linear and sublinear in the
feedforward inputs. 
  In studies of 2-dimensional systems (one excitatory and one inhibitory
population), we will find that, when the excitatory-neuron element of
$-\J^{-1}\g$ (equivalently of $\vr_0$) is negative, the sublinear term becomes dominant (as it
must: $(\y_0)_E<0$, so one must have $\beta(\y_1)_E>|(\y_0)_E|$ for
$y_E>0$) and network behavior becomes strongly sublinear.  In this
case, excitatory firing rates eventually peak and then are ultimately
pushed to zero with increasing $c$, \ie\ with decreasing $\beta$, but
there is typically a large dynamic range of $c$ beyond the
supralinear-to-sublinear transition before this peak occurs (see
Figure~\ref{parameters}).  The
behavior from $c=0$ until somewhat beyond the peak yields behavior
much like that seen in biology, and so we guess that this dynamic
range represents the dynamic range of the feedforward input to cortex.
This will be discussed in Sections \ref{nege}-\ref{solutions}. 
In simulations, the dependence
of $\y_1$ on $\beta$ remains weak, so that $\beta \y_1$ decreases nearly as
fast as $\beta$, as was assumed, over the range from the supralinear-to-sublinear
transition until $r_E$ has been pushed close to zero.
By contrast, when the elements of $-\J^{-1}\g$ are positive, the
subleading correction, $c^{\frac{1}{n}}\r_1$, can have either sign,
and  the total response $\r=c \r_0+ c^{\frac{1}{n}}\r_1$ can be a
supralinear or sublinear function of $c$,\footnote{By saying that
  $\r(c)$ is a supralinear or sublinear function of $c$, we mean that
  $\frac{d^2\r}{dc^2}$ is $>0$ or $<0$ respectively.}
according to whether the sign of $\vr_1$ is negative or positive respectively.  
However, even in the case where the response as a function of $c$
remains supralinear for large $c$ (small $\beta$),  the response to
the sum of two non-aligned inputs, 
$\g_1$ and $\g_2$, will typically sum sublinearly for large $c$, as we
will see in Section \ref{normsec}.
Given positive elements of $-\J^{-1}\g$, Eq.~\ref{y1equ} shows that
$\y_1$ goes to an $O(1)$ 
constant as $\beta$ decreases to zero ($\y_1\rightarrow
\J^{-1}(-\J^{-1}\g)^{.\frac{1}{n}}$), and thus the term $\beta\y_1$
is guaranteed to decay approximately as $\beta$ for sufficiently small $\beta$.

A more systematic account of the large $\alpha$ (small $\beta$) case
can be obtained by formulating a solution like that of Eq.~\ref{yalpha} for
small $\alpha$.  When the elements of $\y$ are $> 0$ (and thus
the elements of $\J\y+\g$ are $> 0$), we can rearrange
Eq.~\ref{yss} for the steady state as
\begin{equation}
\y = -\J^{-1}\g + \beta \J^{-1}\y^{.\frac{1}{n}}
\label{yssbeta}
\end{equation}
Then we can formally write a steady-state solution by starting with
equation \ref{yssbeta} and iteratively substituting $-\J^{-1}\g +
\beta \J^{-1}\y^{.\frac{1}{n}}$ for $\y$ to obtain
\begin{equation}
\y = -\J^{-1}\g + \beta \J^{-1}(-\J^{-1}\g + \beta \J^{-1}(\ldots)^{.\frac{1}{n}})^{.\frac{1}{n}}
\label{highalpha}
\end{equation}
or 
\begin{equation}
\r = -c\W^{-1}\g
+\frac{1}{k^{1/n}}\W^{-1}(-c\W^{-1}\g+\frac{1}{k^{1/n}}\W^{-1} (\ldots)^{.\frac{1}{n}})^{.\frac{1}{n}}
\label{highc}
\end{equation}
If quantities in parentheses are positive, a series solution in powers
of $\beta$ can be obtained from Eq.~\ref{highalpha} in the same manner
as outlined for Eq.~\ref{yalpha}.  When this series converges, which
it will for small enough $\beta$, it gives a steady-state solution.
However, if elements of $-\J^{-1}\g$ are negative, then for small
enough $\beta$ the elements in parentheses will no longer be positive
(and correspondingly, as mentioned above, in the 2-D case $y_E$ is
pushed to zero with decreasing $\beta$ at finite $\beta$, so that
Eq.~\ref{yssbeta} fails at that point).  We can instead regard
Eq.~\ref{yssbeta} as an iterative scheme, $\y[p+1]=-\J^{-1}\g +\beta
\J^{-1}\y[p]^{\frac{1}{n}}$, beginning from some initial condition
$\y[0]$ (Eq.~\ref{yalpha} can also be regarded in this way), which
generates Eq.~\ref{highalpha} as $p\rightarrow\infty$.  Writing
this as $\y[p+1]=f(\y[p])$, if all of the eigenvalues of the Jacobian
of $f$ at the fixed point have absolute values less than 1, then the
iteration will converge to the fixed point within some basin of
attraction about the fixed point.  Hence with suitable initial
conditions, one can find solutions through this iterative scheme,
although not for $\beta$'s less than that at which some elements of
$\y$ are pushed to zero.

These scaling arguments provide key insights into the supralinear
network (Eq.~\ref{requ}) that is confirmed by other analysis and
simulations: for small $\alpha$, recurrence is weak and the network
supralinearly adds responses to different feedforward inputs; with
increasing $\alpha$, there is a transition, for $\alpha=O(1)$, to a
dynamic stabilization that causes the net input neurons receive, and in
many cases their responses, to add sublinearly.  Note
that, when responses add sublinearly, individual neurons still supralinearly sum the net (feedforward
plus recurrent) inputs they receive, but the network ``conspires'' to
deliver net input that is so strongly sublinear that, even after the
neuron raises its net input to the power $n$, its responses add
sublinearly.  We have found in both high-dimensional and 2-dimensional
simulations, and we will show below for the 2-dimensional case, that
stabilization will occur provided feedback inhibition is sufficiently
strong and the inhibitory time constant is not too slow relative to
the excitatory time constant.  This transition from supralinear to
sublinear behavior in turn appears to underly a wide variety of
nonlinearities in neocortical behavior.

\subsection{Comparison to the balanced network}

Van Vreeswijk and Sompolinsky 
\citeyear{vanVreeswijk_Sompolinsky96,vanVreeswijk_Sompolinsky98}
introduced the ``balanced network'' model (see also
\citeA{Renart_etal10}).  They considered a circuit of randomly
connected stochastic
excitatory and inhibitory units that could have activity states 0 or 1,
in the limit in which both feedforward and recurrent
inputs were very large.  (The same model could also be studied in other
regimes, but we use ``balanced network'' to refer to the model in the
large-input regime that they studied and ``balanced state'' to 
refer to the solution they characterized in that regime.)
They studied the conditions in which the network would dynamically find
its way to a balanced state in which the mean input is subthreshold,
yet firing rates are nonzero (where firing rate is defined as the average activity), meaning firing is driven by
fluctuations. They assumed each unit received $K$ inputs of strength
$\frac{1}{\sqrt{K}}$, or a net input of strength $\sqrt{K}$, for $K$
large (\eg, thousands of inputs).  The mean field equations for the
average $E$ and $I$ firing rates are the 2-dimensional version of the
rate equation, Eq.~\ref{rate}, for one E and one I population,
where both $\W$ and $c\g$ are of order
$\sqrt{K}$,\footnote{The input is expressed in units of a variance term which
  itself is dynamically determined, but this term is $O(1)$ and does
  not impact the points made here.}
and the function $f$ is a sigmoidal function rising from 0
to 1 as the input moves from approximately $-3$ to $3$, and saturating
at $0$ or $1$ for smaller or larger values respectively.  To be in the
balanced state, the mean firing rate must be neither 0 nor saturated
at 1, so the net input must be $O(1)$ (\ie\ between $-3$ and 3).  

Thus, the condition for the balanced state is that $\W\r+c\g \sim O(1)$
where both $\W$ and $c\g$ are $O(\sqrt{K})$. The solution, much as in
our scaling argument, is to write $\r=\r_0 + \frac{1}{\sqrt{K}}\r_1 +
\ldots$, where $\r_0$ and $\r_1$ are $O(1)$ and the dots represent
higher-order terms in $\frac{1}{\sqrt{K}}$.  The balance condition is
that the $O(\sqrt{K})$ term in the input vanishes, that is,
$\W\r_0=-c\g$, leaving as input only the $O(1)$ term
$\frac{\W}{\sqrt{K}}\r_1$ and terms that are $O(\frac{1}{\sqrt{K}})$.

The dynamic cancellation condition for the balanced state,
$\r_0=-c\W^{-1}\g$, is of course identical to the condition we have found
for approximate dynamic cancellation in the SSN.\footnote{We wrote the leading
  term as $c\r_0$ rather than $\r_0$, but the leading terms are
  identical.} Although the condition is 
formally identical, the meaning is different in crucial ways:
\begin{enumerate}
\item In the balanced network, the cancellation is required because
  inputs are large and must cancel to leave something small, in order
  to avoid zero or saturated output.  
  In the SSN, partial cancellation
  can already arise when none of the inputs are large and is required so that the
  supralinear input-output functions do not give rise to an
  inconsistent, explosive scaling.  The difference in the size of the
  inputs when cancellation occurs can be seen by recalling that, in
  the SSN, $\alpha=(\text{recurrent weight})(\text{feedforward
    strength})^{n-1}$ (see paragraph above
  Eq.~\ref{yss}).\footnote{Note that this expression for $\alpha$
    holds when the input is scaled appropriately so that 
    the steady state firing rates,
    $f(\text{input})$, are $O(1)$ when the input is $O(1)$ (see
    paragraph above Eq.~\ref{yss}). This relationship between input
    and steady state rates also holds
    for the balanced network, so we have a common scale for the input on which
    to compare the size of the recurrent weights and of the feedforward
    strengths in the two models.}
  In the balanced network, the recurrent weights and the feedforward
  weight are both $O(\sqrt{K})$, so the SSN's $\alpha$ would be
  $O(K^{\frac{n}{2}})$. 
  In contrast, dynamic cancellation arises in the SSN when $\alpha$
  is $O(1)$.
\item In the balanced network, the second-order term
  $\frac{1}{\sqrt{K}}\r_1$ is negligibly small relative to the
  first-order term $\r_0$ (because the stabilization is to cancel
  large things, leaving something small).  The first-order term is
  linear in the input, $\r_0=-c\W^{-1}\g$, and so in the balanced
  network responses are always linear in the input.  
  In the SSN, the first-order term in $\y$, $y_0=J^{-1}\g$, is
  $O(1)$, and while the second-order term $\beta y_1$ scales approximately
  as $\beta$, it can have a large pre-factor such that it
can be comparable to or larger than the first-order term over a wide
dynamic range, enabling a variety of sublinear behavior (and 
$\r_0$ and $\r_1$ are just $c/\psi$ times the corresponding $\y$'s).  

  In particular, in the SSN, elements of $\r_0$ can be negative,
  meaning that for such an element $r_1>|r_0|$ over the relevant dynamic
  range of behavior (discussed in more detail for the 2-D model in
  Section \ref{nege}).  In the balanced network, since all terms except
  $\r_0$ are negligibly small, the elements of $\r_0$ must be positive for
  activity to be nonzero.
\end{enumerate}

In sum, in the balanced network, inputs are huge relative to the
distance from rest to threshold, and must dynamically cancel for the
network to neither saturate nor have 0 activity but instead 
be in the fluctuation-driven regime. The dynamic cancellation or 
stabilization yields network responses that are always linear in the
input.  In the SSN, the supralinear input-output function renders
the network explosive -- input is raised to a power greater than 1 to
produce responses, which feed back as input.  Stabilization
against this explosive nonlinearity arises when inputs are relatively small,
yielding a range of sublinear behavior.

Finally, for clarity we note that the balanced network and SSN are not
necessarily different models, but rather very different solutions with
different behaviors in different regimes, which may be found in the
same model. For example, in the power-law model studied here, the
balanced-state solution can be found if $\alpha$ is large, the
elements of $\y_0=-\J^{-1}\g$ are positive, and the feedforward input
$k^{\frac{1}{n}}c$ and recurrent input $k^{\frac{1}{n}}\psi$ both have
the same scaling, \eg\ both 
are $O(\sqrt{K})$, so that $(c/\psi)$ is $O(1)$.  In this case, as
$\alpha\rightarrow\infty$, $\y\rightarrow \y_0$ and
$\r\rightarrow \frac{c}{\psi}\y_0$, so $\r$ is $O(1)$; and the
first-order correction is $c^{\frac{1}{n}}\r_1=\frac{c}{\psi}\beta \y_1 =
\left(\frac{c}{k\psi}\right)^{1/n}\frac{1}{\psi} \y_1$, which scales as
$\frac{1}{k^{\frac{1}{n}}\psi}$, \eg\ as $\frac{1}{\sqrt{K}}$ (recall
$\y_1$ is $O(1)$ for large $\alpha$). This is the balanced-state solution.
Similarly, in the model studied in the balanced
network, the input/output function is initially supralinear before
saturating, so we expect that SSN-like solutions could be found for
appropriate regimes of input and recurrent strengths with
$\alpha=O(1)$.\footnote{The value of $n$ in the definition of $\alpha$
  could be derived from an approximate power-law fit to the supralinear
  portion of the input/output function in that model.}

\section{Reduction to a 2-dimensional system}

Most of our analysis hereafter will focus on a 2-dimensional system of
one excitatory and one inhibitory population, as it is difficult to
say much in general in higher dimensions. A 2-D system of one E and
one I population can be derived as a mean field equation from
higher-dimensional models in which E and I neurons have random
connectivity
(\eg\ \citeA{vanVreeswijk_Sompolinsky98}, \citeA{Renart_etal10}). In
particular, if the high-dimensional model involves integrate-and-fire
neurons, their input-output functions in the fluctuation-driven regime
can be reasonably approximated by power-law functions
\cite{Hansel_vanVreeswijk02}.

Here we consider a higher-dimensional system with structured
connectivity.  We show a heuristic derivation of a 2-D system that
preserves a surprising amount of the behavior of the
higher-dimensional system.  We then show how the conditions for
``normalization'' -- sublinear addition of responses to multiple
stimuli -- in the high-D system can be expressed as simple conditions
in the 2-D system on the growth of $\r$ with increasing $c$, or the
growth of $\y$ with increasing $\alpha$.  In particular, the
transition to the high-$\alpha$ regime represents a transition to
sublinear addition in the corresponding high-D system, even when the
dependence of $\r$ on $c$ remains supralinear in the 2-D system.

\subsection{Reduction}
\label{sec:reduction}

We consider a topographic network, with pairs of excitatory ($E$) and
inhibitory ($I$) units arranged on a 1-D or 2-D grid with periodic
boundary conditions. The grid dimensions mirror stimulus parameters
such as orientation or position on the retina, such that units at a
certain location in the grid prefer stimuli with the corresponding
parameter(s) (other stimulus preferences may also be incorporated, \eg, a 2-D retinotopic grid with a superimposed map of preferred
orientations). We will use as an illustration a 1-D ring of cells of
similar retinotopic position but varying preferred orientation, with
preferred orientation represented by position on the ring, but the
reduction framework is more general.  Stimuli are localized on the
grid (\eg\ in the 1-D ring a single
oriented grating stimulus evokes a localized Gaussian-shaped  bump of input,
centered at the neurons that prefer the stimulus orientation), though there 
may be more than one localized stimulus present (\eg\ two superimposed
oriented gratings of different orientations). For simplicity, we 
assume a single time constant for all $E$ cells and one for all $I$ cells.
We let $\theta$ represent the position on the grid, and $r_E(\theta)$
and $r_I(\theta)$ the excitatory and inhibitory firing rates at
position $\theta$.  
Thus, we can write Eq.~\ref{requ} as
\begin{eqnarray}
\label{ringE}
\tau_{E} \frac{d r_{E}(\theta)}{dt} &=& -r_{E}(\theta) + k\left(W_{EE}
\ast r_{E}(\theta) - W_{EI} \ast r_{I}(\theta) +  c g_{E}(\theta)
\right)^{.n} \\
\tau_{I} \frac{d r_{I}(\theta)}{dt} &=& -r_{I}(\theta) +
k\left(W_{IE}
\ast r_{E}(\theta) - W_{II} \ast r_{I}(\theta) + c g_{I}(\theta)\right)^{.n}
\label{ringI}
\end{eqnarray}
Here, $W_{XY}\ast r_Y(\theta)=\sum_{\theta'}W_{XY}(\theta,\theta')r_Y(\theta') \Delta \theta$, where $\Delta \theta$ is the stimulus parameter volume per grid point.

We will consider ``normalization", the sublinear addition of the
responses to two stimuli \cite{Carandini_Heeger11}.  We let one
stimulus be centered at 
$\theta=0$.  We  let $\tilde W_{XY}=W_{XY}(0,0)$, and we define $\vec w_{XY}\equiv \frac{W_{XY}(0,\theta)}{W_{XY}(0,0)}\Delta\theta$ to be the vector of weights to position 0, normalized by $\tilde W_{XY}$.  Similarly, we let $\hat\r_{E}$,
$\hat\r_{I}$ be the vectors of excitatory and inhibitory firing rates,
respectively, normalized to equal 1 at position 0, with elements $\hat
r_{E}(\theta)$, $\hat r_{I}(\theta)$.    Then the equations
for the units at position 0 are
\begin{eqnarray}
\label{eqn18}
\tau_{E} \frac{d r_{E}(0)}{dt}&=& -r_{E}(0) +k\left(
 \tilde W_{EE} r_E(0)\left( \vec{w}_{EE} \cdot \hat\r_{E} \right)  -
 \tilde W_{EI} r_I(0)\left( \vec{w}_{EI} \cdot \hat \r_{I} \right)  +
c g_{E}(0)\right)^{.n} \label{highde}\\ 
\tau_{I} \frac{d r_{I}(0)}{dt}&=& -r_{I}(0) +k\left(
\tilde W_{IE} r_E(0)\left( \vec{w}_{IE} \cdot \hat\r_{E} \right)  - 
 \tilde W_{II} r_I(0)\left( \vec{w}_{II} \cdot \hat \r_{I} \right) 
+ c g_I(0)\right)^{.n} \label{highdi}
\end{eqnarray}
Although we had previously incorporated changes in $|\g|$ into $c$, we
now take addition of a second stimulus to
simply alter $\g$ with no change in $c$, so that in particular
addition of a second stimulus that gives no input to position 0 does
not alter $g_E(0)$ or $g_I(0)$.  We now define
\begin{eqnarray}\label{jxydef}
J_{XY}&\equiv&\tilde W_{XY}
\frac{ \left( \vec{w}_{XY} \cdot \hat \r_{Y}\right) } { \left(
    \vec{w}_{EE} \cdot \hat \r_{E}\right) }\\
\Psi &\equiv&
 \vec{w}_{EE} \cdot \hat \r_{E}
\label{psidef}
\end{eqnarray}
Note that in general the $J_{XY}$ depend on the normalized shapes of the
responses, $\hat r_E$ and in some cases $\hat r_I$, which in turn may
depend on $c$ and/or $\g$.  Letting $r_E\equiv r_E(0)$, $g_E\equiv
g_E(0)$, etc., our equations become
\begin{eqnarray}
\tau_E \frac{d r_E}{dt} &=& -r_E + k\left(\Psi J_{EE} r_{E} -
\Psi J_{EI} r_{I} + c g_{E}\right)^{.n} \label{twode}\\ 
\tau_I \frac{d r_I}{dt} &=& -r_I + k\left(\Psi J_{IE} r_{E} -
  \Psi J_{II} r_{I} + c g_{I}\right)^{.n}
\label{twodi}
\end{eqnarray}
As written, Eqs.~(\ref{twode})--(\ref{twodi}), together with the
definitions (\ref{jxydef}) and (\ref{psidef}), are simply a rewriting
of Eqs.~(\ref{highde})--(\ref{highdi}) and hence exact.  In
particular, the seemingly closed system of equations for the two
variables $r_E = r_E(0)$ and $r_I = r_I(0)$ are in general
parametrically dependent on the values of $\hat{r}_X(\theta)$ at other
$\theta$'s through the dependence of $J_{XY}$ and $\Psi$ on the
normalized shapes of the response curves. 

We now adopt the ansatz that, as the stimulus changes, the four dot 
products $\vec{w}_{XY} \cdot \hat \r_{Y}$ are all scaled by a common
factor, whether the stimulus changes in strength (changing $c$) or in
shape (changing $\g$, \eg\ by adding a second stimulus). This means that
we treat the $J_{XY}$ as constants independent of $c$ and $\g$. On the other hand, $\Psi$ is
scaled by this common scaling factor.  We make the further ansatz that
$\Psi$ depends only on $\g$ and not on $c$. Since 
$\Psi$ depends only on the shape of the response,
$\hat{\r}_E$, and not its magnitude, this amounts to the
ansatz that stimulus strength $c$ alters
response magnitude without altering response shape.
With these two ansatze,
Eq.~\ref{twodi} is simply the 2-dimensional version of Eq.~\ref{requ},
and is equivalent to Eq.~\ref{yequ} for a 2-dimensional $\y$ if $\Psi$ replaces $\psi$ in the definitions of $\y$
and $\alpha$ (Eq.~\ref{dimensionless}). (We are no longer following
  our convention of $\|\J\|=1$, as 
the matrix $\J$ composed of these $J$'s need not satisfy $\|\J\|=1$
for any standard matrix norm; one could, however, return to this
convention by dividing the
$J_{XY}$'s by $\|\J\|$ and multiplying $\Psi$ by the same factor.)  If,
furthermore, we also assume that the shapes 
of the population responses roughly follow the shape of the input, then,
since the weights
$\vec{w}_{XY}$ are non-negative, the effect of adding a second
non-negative stimulus is to increase $\Psi$. Hence ``normalization''
of the E or I population corresponds to a decrease in firing rates
$r_E(0)$ or $r_I(0)$, respectively, with increasing $\Psi$, which for a second stimulus of infinitesimal strength corresponds to 
\be
\label{normcondition}
\frac{d
  r_E(0)}{d\Psi}<0, \qquad \frac{d r_I(0)}{d\Psi}<0.
\ee

The ansatze, of course, are not in general true, but they can be close
enough to true to give a good qualitative account of the higher-dimensional
system (however, see the discussion at the end of Section~\ref{solutions}
for a discussion of cases where the assumptions in our ansatze fail considerably and
therefore \req {normcondition} no longer expresses correctly the
condition for normalization).  To illustrate this, we simulate the
model on a 
one-dimensional ring, which we think of as representing the preferred
orientations of neurons representing a  common position in visual space.\footnote{The paradigm we study here -- suppression of
  response to one orientation by presentation of an orthogonal
  orientation -- is known as ``cross-orientation suppression''. In V1,
  this appears to be primarily mediated by sublinear addition of the
  feedforward inputs to V1 evoked by the two stimuli
  \cite{Lauritzen_etal01,Priebe_Ferster06,Li_etal06c}. However we use
  this paradigm to study how the model cortex sums responses to
  multiple stimuli, assuming the feedforward inputs sum linearly.} We
consider 180 E/I pairs at grid positions separated by $1^{\mathrm{o}}$ in
preferred orientation, with $0^{\mathrm{o}}=180^{\mathrm{o}}$.  All four connection types
have the same width, following evidence that excitatory and inhibitory
inputs received by cells in upper layers have similar orientation
tuning
\cite{Ferster86,Anderson_etal00,Martinez_etal02,Marino_etal05}.  The
connectivity takes the form
\begin{equation}
W_{XY}(\theta,\theta')=J_{XY}e^{-\frac{d(\theta,\theta')^{2}}{2\sigma_{\msub{ori}}^{2}}}
\label{wequ}
\end{equation}
where $d(\theta,\theta')$ is the shortest distance around the circle
between $\theta$ and $\theta'$.  We
consider stimulation by either one oriented luminance grating or two
orthogonal gratings of equal contrast. Each grating is represented by a Gaussian-shaped
curve of feedforward input with width (standard deviation of the
Gaussian) $\sigma_{\mbox{stim}}$
and height $c$; a single grating is centered at
$\theta=0^{\mathrm{o}}$, a second added grating is centered at $\theta=90^{\mathrm{o}}$. For any given stimulus (1 or 2
stimuli, stimulus height $c$, given stimulus width
$\sigma_{\mbox{stim}}$) the equivalent 2-D  
model is found as follows: we use the same $\tau_{E,I}$, $k$, $n$, and
$J_{XY}$'s as in the high-dimensional model
and take $\Psi$ to be the value of the convolution, at $\theta=0$, of
the connectivity Gaussian (Eq.~\ref{wequ} with $J_{XY}=1$, \ie\ the
vector of weights to $\theta=0$ normalized to equal 1 at $\theta=0$) with
$\g^{.n}$ (for our stimuli, $\g$ is 1 at position 0). We use $\g^{.n}$ 
 as a surrogate for the shape of
the response, with the knowledge that
at least at low contrast it gives a good approximation to this shape
(Section~\ref{subsec:smallalpha}). 

\begin{figure}[htp]
\centering
\vspace{-1 cm}
\includegraphics[width=\textwidth]{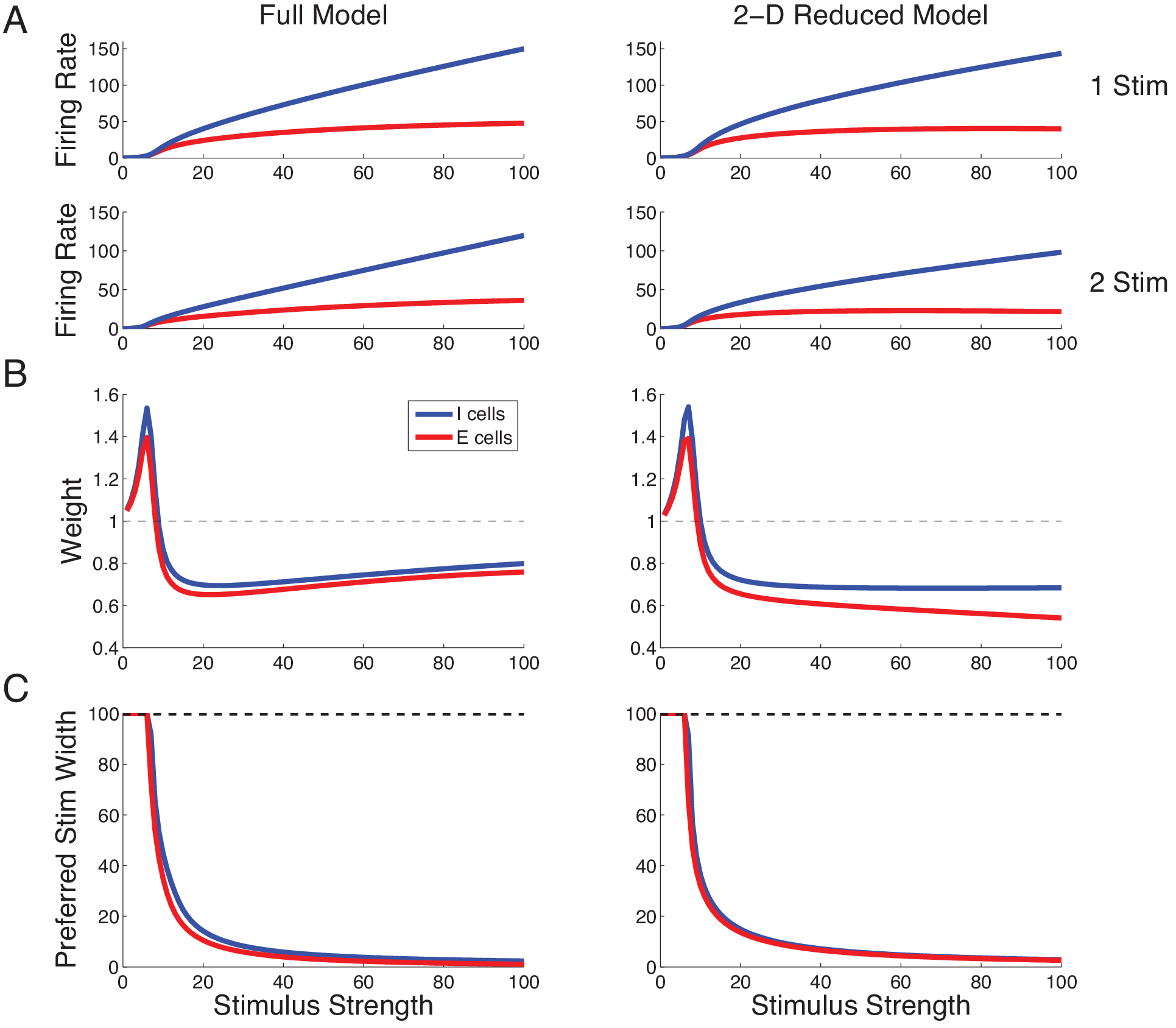}
\vspace{ 0 cm}
\put(-444,226){\rotatebox{90}{{$w$}}}
\put(-400,111){${\scriptstyle \infty}$}
\put(-408,118){\rotatebox{135}{\textbf{$\to$}}}
\put(-185,111){${\scriptstyle \infty}$}
\put (-193,118){\rotatebox{135}{\textbf{$\to$}}}
\caption[Two neuron approximation of the full ring model.]
{\small
  {\bf Two neuron approximation of the full ring model.}  
  \rm~(A) The reduced version
  of the model (right) produces qualitatively similar curves of response vs.\
  stimulus strength $c$ as the full model (left; for the full model, this is the response of the
  cells at $\theta=0^\deg$). The top plots show the response curves for a
  stimulus composed of a single grating with orientation $\theta=0^\deg$
  and the bottom 
  plots show the response for a two-grating stimulus composed of 
  the grating at $\theta=0^\deg$ and a grating at $\theta=90^\deg$.
 In the 2-D reduced model,  
 these two cases are represented by using $\Psi=0.774$ for one grating
 and $\Psi=1.024$ for two gratings (see \req {psidef} for the definition of $\Psi$ and the text after \req {wequ}  for the method we used to calculate these values).
 (B) Full and reduced models show a similar
  stimulus-strength-dependent transition from supralinear summation
  (weight $>1$) to sublinear summation (weight $<1$) of the responses
  to two gratings, where the weight $w$ is defined as follows. For the full model, for either E or I cells, we let
  $R_1(\theta)$, $R_2(\theta)$, and $R_{12}(\theta)$ be the response
  to one grating, the other grating, or the superposition of the two,
and we define
  $w = \frac{R_{12} (0)}{R_1 (0) +R_2 (0)}$, where $\theta = 0$ is the orientation of the first grating.  For the
  reduced model, we define the weight as $w=\frac{R_{12}}{R_1 + R_2}$, where $R_1$, and
  $R_{12}$ are the responses to one or two gratings (modeled by the two values of $\Psi$ given above) and we set $R_2 =0$  (by the way we defined the reduction, $R_1$, $R_2$ and $R_{12}$ should approximate the responses of the full model at $\theta=0$).  (Continued on next page)
  \smallskip}
\label{analyticRing}
\end{figure}
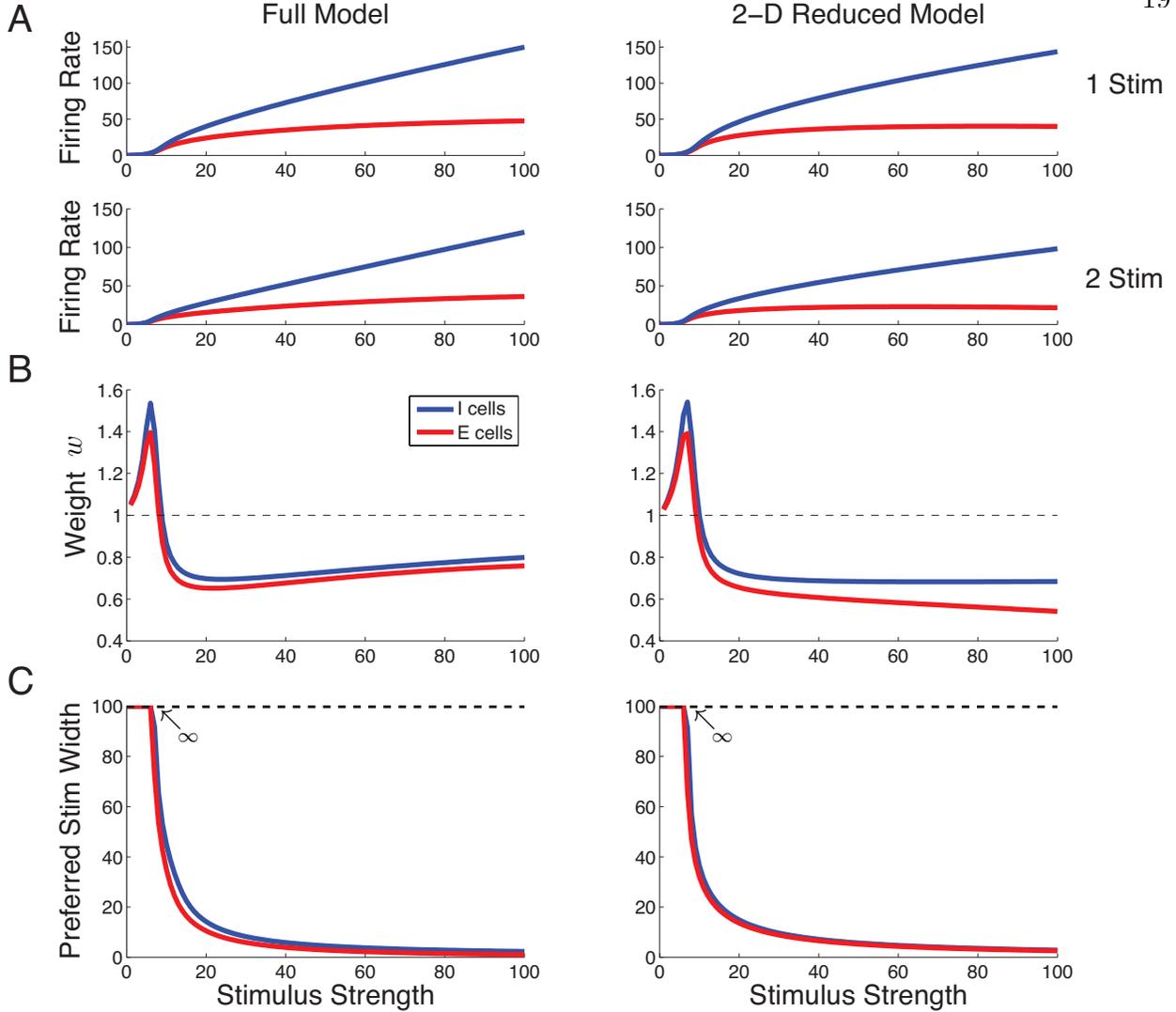

\begin{figure}
\contcaption{\small (Continued). 
(C) Full
  and reduced models have nearly identical stimulus-strength-dependent
  tuning for the width in orientation, $\sigma_{\mbox{stim}}$, of a
  feedforward stimulus (full model: width of 
  Gaussian stimulus centered at $\theta=0$ with given stimulus
  strength $c$ that gives the strongest response in cells at
  $\theta=0$; reduced model: $\Psi$ is computed for each stimulus
  width as described in the text after \req {wequ}, and plot shows width whose $\Psi$ gives maximal response). In all curves, red shows E cells and blue shows I
  cells.  All responses are steady-state responses.  Full model
  solutions found by simulating until convergence to steady state.
  Parameters: $J_{EE} = 2.5$, $J_{IE} = 2.4$, $J_{EI} = 1.3$, $J_{II}
  = 1.0$, $\tau_{E} = 20~{\mbox
  ms}$, $\tau_{I} = 10~{\mbox ms}$, $k = 0.04$, $n = 2.0$, 
$\sigma_{\msub{ori}}=32^{\mathrm{o}}$; $\sigma_{\mbox{stim}}=30^\deg$
in A,B.
    \smallskip}
\end{figure}

The result is that the reduced 2-D model accurately reproduces the
behavior of the full model as shown in Fig.~\ref{analyticRing}. {(See also Figs.~\ref{parameters}--\ref{parameters-Nd} of Section~\ref{solutions} for a more detailed comparison in various parameter regimes; as explained at the end of Section~\ref{solutions},  the bottom three rows of Figs.~\ref{parameters-Nd} show directly the quality of the approximations involved in the ansatze introduced after \req {twodi}.)}  The firing
rates of the cells at $\theta=0$ vs.\ stimulus strength closely match
the firing rates in the 2-D model (Fig.~\ref{analyticRing}A).  Both
models show a similar transition from supralinear summation of
responses to the two gratings for weak stimuli to sublinear summation
or ``normalization'' for stronger stimuli (Fig.~\ref{analyticRing}B).
The network also shows a form of surround suppression, in which the
``summation field size'' -- the stimulus width 
that yields maximal response for a given stimulus strength -- shrinks
monotonically with increasing stimulus strength, as is well known in
real space (rather than orientation space) for V1 cells
\cite{Sceniak_etal99,Cavanaugh_etal02a}, and this behavior is
extremely similar in the full and reduced models
(Fig.~\ref{analyticRing}C).\footnote{Note that this ``summation field
  size'' for orientation selectivity should not be confused with the
  orientation tuning width, which is the width of the orientation
  tuning curve obtained by studying response vs.\ single orientations
  (more precisely: studying response vs.\ center orientation, using
  stimuli that evoke a fixed curve of feedforward input
  vs. orientation that is symmetric about the center orientation). The
  orientation tuning curve, representing the set of single
  orientations that can drive the cell, is analogous to the ``minimal response
  field'' in real space, which represents the sum of the set of small
  regions in visual space in which appropriate light stimuli can
  evoke spiking responses. The minimal response field in real space is
  invariant with stimulus contrast \cite{Song_Li08}, and so too is the shape
  of the orientation tuning curve 
  \cite{Skottun_etal87,Anderson_etal00b,Ferster_Miller00} (contrast is
  monotonically related to the firing rate of the inputs to cortex
 (\eg\  \citeA{Ohzawa_etal85}).  The fact
  that the summation field size in real space is larger than the
  minimal response field indicates that stimuli in regions where light
  cannot directly drive spikes can facilitate responses to stimuli
  in the minimal response field. Recall that the size of this facilitating area
  shrinks with contrast.  The model suggests that the same may be true
  in the orientation domain, in terms of cortical processing of
  feedforward input to cells of different preferred
  orientations. However, attempts to test this idea will likely be compromised
by two facts: (1) simultaneous presentation of multiple orientations
does not yield linear summation of the input to cortex evoked by the individual
orientations \cite{Lauritzen_etal01,Priebe_Ferster06,Li_etal06c} and
(2) varying the feedforward orientation tuning by changing stimulus
attributes -- \eg\ a sinusoidal luminance grating of a given size provides
drive to cortical cells with an orientation tuning that 
narrows with increasing spatial frequency, and similarly a longer bar
drives narrower orientation tuning than a shorter bar -- also changes
other attributes to which the neurons are independently sensitive,
such as spatial frequency or bar length.}
Thus, the 2D model can provide a good basis for understanding more
general models.

\subsection{Conditions for Normalization in the 2-Dimensional System}
\label{normsec}

Here we show that, when our ansatze hold, the high-dimensional network
exhibits normalization precisely when the 2-dimensional network shows
sublinear scaling.
We consider steady-state $\r$ or $\y$ and use expressions like
$\frac{\partial r_X}{\partial \psi}$ to refer to the dependence of the steady state on
parameters.  We have seen that $r_X$ ($X\in\{E,I\}$) exhibits normalization in response to addition of an infinitesimal second stimulus if
$\frac{\partial r_X}{\partial \Psi}<0$ in the 2-D model (more generally, for a finite-strength second stimulus, if $\int_{\Psi_{\msub{init}}}^{\Psi_{\msub{final}}}
 \frac{\partial r_X}{\partial \Psi} d\Psi <0$).
Since Eqs.~\ref{twode}-\ref{twodi} are equivalent to
Eqs.~\ref{dimensionless}-\ref{yequ} with $\Psi$ replacing $\psi$, we
revert to the notation of Eqs.~\ref{requ}-\ref{yequ} and use $\psi$.

We work with the 2-D model and express the conditions
$\frac{\partial r_X}{\partial \psi}<0$ as a single vector condition.
We note first that
$\frac{\partial \r}{\partial \psi}=
c\frac{d\y/\psi}{d\psi}=c\left(\frac{1}{\psi}\frac{d\y}{d\psi}-\frac{\y}{\psi^2}\right)$
and
$\frac{d\y}{d\psi}=\frac{d\y}{d\alpha}\frac{d\alpha}{d\psi}=k
c^{n-1}\frac{d\y}{d\alpha}$. Putting these together we find
$\frac{\partial \r}{\partial \psi}=
\frac{k c^n}{\psi} \left( \frac{d\y}{d\alpha} - \frac{\y}{\alpha}\right)$. 
Thus, the condition for normalization is that $\y$ grow more slowly
than linearly with increasing $\alpha$:
$ \frac{d\y}{d\alpha} <
\frac{\y}{\alpha}$ or $\frac{d\ln\y}{d \ln \alpha}<1$ or, roughly,
that $\y\sim \alpha^p$ for $p<1$.  As we have seen, $p$ becomes less
than $1$ precisely when the transition from the supralinear to the sublinear
scaling regime occurs.

We can reexpress this in terms of $\r$.  Using algebra similar to the
above, we find $\frac{\partial \r}{\partial c}=\frac{(n-1)\alpha}{\psi}\left(
  \frac{d\y}{d\alpha} + \frac{\y}{(n-1)\alpha}\right)$, from which we
find that $ \frac{d\y}{d\alpha} < \frac{\y}{\alpha}$ is equivalent to
$\frac{\partial \r}{\partial c}<n\frac{\r}{c}$.  Thus, the condition for normalization
is that $\r$ grow more slowly than $c^n$ with increasing $c$:
$\frac{d\ln\r}{d \ln c}<n$ or, roughly, that $\r\sim c^p$ for
$p<n$.  Again, $p$ becomes less than $n$ precisely at the transition
from supralinear to sublinear scaling.

Finally, noting that the steady state condition is
$\r=k(\W\r+c\g)^{.n}$, without loss of generality we write $\W\r = c
\f(c)$ for some vector function $\f$ of $c$, so that the steady state
condition $\r=k(\W\r+c\g)^{.n}$ becomes $\r=k c^n(\f(c)+\g)^{.n}$.  Thus
we see that a component of $\r$ grows more slowly than $c^n$ precisely
when the corresponding component of $\f(c)$ is a decreasing function
of $c$ (that is, for corresponding components $r$ and $f$,
$\frac{\partial r}{\partial c}<n\frac{r}{c}$ precisely when $f'(c)<0$).  Thus, the
condition for normalization can alternatively be expressed as the
requirement that the recurrent input $\W\r$ grow more slowly than
linearly with $c$, \ie\ 
$\frac{\partial(\W\r/c)}{\partial c}<0$ yields
$\frac{\partial\W\r}{\partial c}<\frac{\W\r}{c}$ or
$\frac{\partial\ln(\W\r)}{\partial\ln c}<1$.

\section{Analyses of the 2-Dimensional Network}

\def\nege{\Omega_E}
\def\negi{\Omega_I}
We will assume throughout this analysis that $g_E\geq 0$, $g_I\geq 0$.
We will use the following definitions:
\begin{eqnarray}
\label{Omegadef}
\nege &\equiv& \Det\, \J \left(-\J^{-1}\g\right)_E = J_{II}g_E-J_{EI}g_I\\
\negi &\equiv& \Det\, \J \left(-\J^{-1}\g\right)_I = J_{IE}g_E-J_{EE}g_I
\end{eqnarray}
We also note that there are three possible conditions: (1)
$\left(-\J^{-1}\g\right)_E>0$ and $\left(-\J^{-1}\g\right)_I>0$; (2)
$\left(-\J^{-1}\g\right)_E<0$ and $\left(-\J^{-1}\g\right)_I>0$; and (3)
$\left(-\J^{-1}\g\right)_E<0$ and $\left(-\J^{-1}\g\right)_I<0$.  The
$4^{\msub{th}}$ condition, 
$\left(-\J^{-1}\g\right)_E>0$ and $\left(-\J^{-1}\g\right)_I<0$, is
not mathematically possible for $g_E\geq 0$ and $g_I \geq 0$: 
$\nege>0$ and $\negi<0$ together imply $\Det\, \J<0$, and similarly
$\nege<0$ and $\negi>0$ together imply $\Det\, \J>0$.

\subsection{When Does the Network Dynamically Stabilize?}

\subsubsection{The case of infinitely fast inhibition}

We first analyze the case of infinitely fast inhibition,
$\tau_I/\tau_E = 0$, with constant feedforward inputs. We show that in
this case, if $\Det\, \J>0$, the network is always driven
to a stable fixed point from arbitrary starting conditions.  The
condition $\Det\, \J>0$ means that feedback inhibition is sufficiently
strong: $J_{EI}J_{IE}>J_{EE}J_{II}$. In addition, we show that if
$\Det\, \J<0$, sufficiently large initial firing rates will cause the
system to ``blow up'', \ie\ firing rates will grow arbitrarily large.

With $\tau_I=0$, the value of $y_I$ is ``slaved'' to, or
instantaneously set by the value of, $y_E$ according to the
$\frac{dy_I}{dt}$ part of Eq.~\ref{yequ} for $\y$.  Because of the
nonlinearity, we cannot solve this for $y_I$ as a function of $y_E$,
but we can instead solve for $y_E$ as a function of $y_I$:
\begin{equation}
y_E =
\frac{1}{J_{IE}}\left(
\left(\frac{y_I}{\alpha}\right)^{\frac{1}{n}}+J_{II}y_I-g_I \right)
\label{yeequ}
\end{equation}
Substituting this in the $\frac{dy_E}{dt}$
part of Eq.~\ref{yequ} yields, after a bit of algebra, an equation for
$\frac{dy_I}{dt}$ induced by the slaving of $y_I$ to the $y_E$ dynamics:
\begin{equation}
\tau_E\frac{dy_I}{dt} =
\frac{ n\alpha^{\frac{1}{n}}y_I^{\frac{n-1}{n}} } 
{ 1 + J_{II} n\alpha^{\frac{1}{n}}y_I^{\frac{n-1}{n}} }
\left(
-J_{II}y_I - 
\left( \frac{y_I}{\alpha} \right)^{\frac{1}{n}} 
+g_I + \frac{\alpha}{J_{IE}^{n-1}} 
\left ( -\Det\, \J\, y_I + J_{EE}\left(\frac{y_I}{\alpha}\right)^{\frac{1}{n}}
+\negi \right)^{.n} \right)
\end{equation}

For sufficiently large $y_I$, if $\Det\, \J>0$, the term inside the
parentheses in the $\frac{\alpha}{J_{IE}^{n-1}} \left ( \ldots
\right)^{.n}$ term will be negative, and so will be set to zero after
the thresholding involved in the $()^{.n}$ operation.  The dominant
term will then be the $-J_{II}y_I$ term, which is negative.  So for
sufficiently large $y_I$, $\frac{dy_I}{dt}<0$.  On the other hand, 
if $\Det\, \J<0$, then for sufficiently large $y_I$, the $\left ( \ldots
\right)^{.n}$ will be positive and larger than the sum of the other
terms, so that $\frac{dy_I}{dt}>0$ and, since increasing $y_I$ will
increase $\frac{dy_I}{dt}>0$, this derivative is ever-increasing.

For sufficiently small $y_I$, $\frac{dy_I}{dt}>0$ if either $g_I$ or
$g_E$ is nonzero, which can be seen as follows.  For sufficiently
small $y_I$, the source terms $g_I$ and $\Omega_I$, if nonzero,
dominate the terms involving $y_I$.  Both $g_I$ and the $\left (
  \ldots \right)^{.n}$ term containing $\Omega_I$ are non-negative, so
if either is positive $\frac{dy_I}{dt}$ will be positive; if
$\Omega_I>0$, the $\left ( \ldots \right)^{.n}$ term is positive; if
$\Omega_I\leq 0$, this implies $g_I>0$ (given that at least one of
$g_I$ and $g_E$ is nonzero, and that both are non-negative).

Thus, for $\Det\, \J>0$, $y_I$ is driven to a stable fixed point, and
$y_E$ is then determined from Eq.~\ref{yeequ}, so the system will
arrive at a stable fixed point.  Note that the system could have
multiple fixed points with varying levels of $y_I$.  The topology of
flow along the $y_I$ axis tells us that there must be an odd number of
fixed points, alternating from stable to unstable to stable with
increasing $y_I$, with the outermost fixed points (those with lowest
and highest $y_I$) being stable.  In the simplest case, there is a
single stable fixed point.  In addition, for $\Det\, \J<0$, the system
will blow up for sufficiently large initial firing rates.

\subsubsection{More general requirements for stability}
\label{sec:genreqr}

Changes in the time constants can alter the stability of the fixed
points, but do not alter the number or positions of the fixed points.
The results of the previous section tells us that, for $\Det\,
\J>0$, the system always has a fixed point that is stable for
$\tau_I=0$. We consider such a fixed point, and ask when it retains or
loses stability for finite $\tau_I$.

We let the fixed point be $\vectwo{y_E}{y_I}$, and assess stability by
linearizing the dynamics about this fixed point. We let $q =
\tau_I/\tau_E>0$.  Setting $\tau=\tau_E$ in Eq.~\ref{yequ}, the matrix
$\T$ is given by $\T=\mat{1}{0}{0}{q}$. Define the matrix 
$ 
\bPhiy=n
\alpha^{\frac{1}{n}} \mat{y_E^{\frac{n-1}{n}}}{0}{0}{y_I^{\frac{n-1}{n}}}.
$
Writing the identity matrix as $\1$, the
Jacobian matrix of the 2-D system is:
\begin{equation}
\label{2djacob}
{\cal J}\equiv \mat{{\cal J}_{EE}}{-{\cal J}_{EI}}{{\cal J}_{IE}}{-{\cal J}_{II}}
=\T^{-1}\left(\bPhiy\J-\1\right)
\end{equation}
A fixed-point of the dynamics will be stable if ${\cal J}$ has a
negative trace and a positive determinant.
  
The negative trace condition is ${\cal J}_{EE}<{\cal J}_{II}$, which
becomes 
\begin{equation}
{\cal J}_{EE}=n \alpha^{\frac{1}{n}}y_E^{\frac{n-1}{n}}J_{EE}  - 1\leq
0\,\,\,\text{OR}\,\,\,\left({\cal J}_{EE}>0\,\,\,\text{AND}\,\,\,
q 
< 
\frac{n \alpha^{\frac{1}{n}}y_I^{\frac{n-1}{n}}J_{II}+1}
{n \alpha^{\frac{1}{n}}y_E^{\frac{n-1}{n}}J_{EE}  - 1} \right)
\label{stab}
\end{equation}
The condition ${\cal J}_{EE}\leq 0$
means that the excitatory subnetwork by itself is stable (or
marginally stable),
which guarantees that the network will always be stable, since 
only  unchecked recurrent self-excitation can destabilize the network.  When
the excitatory subnetwork is unstable, we can further reduce the condition
on $q$ for $n=2$:\footnote{This condition \label{footcond} is found by solving 
$\sqrt{y_{I}} = \sqrt{\alpha}\left(J_{IE}y_E - 
  J_{II}y_{I} + g_{I}\right)$ as a quadratic equation for
$\sqrt{y_{I}}$. Discarding the negative 
solution, this yields $\sqrt{y_{I}} = \frac{-1+\sqrt{1+4 \alpha
    J_{II}\left(g_I+J_{IE}y_E\right)}}{2J_{II}\sqrt{\alpha}}$. Substituting
this into Eq.~\ref{stab} for $n=2$ yields Eq.~\ref{newq}.}
\begin{equation}
q<\frac{\sqrt{1 + 4\alpha J_{II}\left(g_I+J_{IE}y_E\right)}}
{2 J_{EE} \sqrt{\alpha y_E } - 1} \qquad 
\left(n=2,\,\,\, 2\sqrt{\alpha y_E}J_{EE}> 1\right)
\label{newq}
\end{equation}

The determinant condition, $\Det\, {\cal J}>0$, is always true for any
fixed point that is stable at $q=0$.  To see this, note that the sign
of the determinant does not depend on $q$ for $q>0$ (because $\Det\,
\A\B=\Det\, \A\,\Det\, \B$ for any matrices $\A,\B$, and $\Det\,
\T^{-1}=\frac{1}{q}$).  So if we prove that $\Det\, {\cal J}>0$ for
arbitrarily small $q>0$, we will have shown that it holds for all
$q>0$. For $q=0$, the determinant, which is the product of the two
eigenvalues, was infinite: because the fixed point was stable, both
eigenvalues had negative real part: one real part was infinite,
corresponding to the instantaneous flow onto the inhibitory nullcline
(the line in the $y_E$/$y_I$ plane on which $\frac{dy_I}{dt}=0$); the
other was finite, corresponding to the flow along the nullcline
converging onto the fixed point.  (Since the two real parts were
unequal, both eigenvalues were real.)  As $q$ is moved infinitesimally
from 0, the infinite eigenvalue becomes a large but finite negative
eigenvalue, while the finite eigenvalue is perturbed by arbitrarily
small amounts as $q$ is made arbitrarily small.  This means that there
is a range of $q>0$ for which the eigenvalues continue to have
negative real parts, and therefore for which the determinant condition
holds.  Therefore, the determinant condition holds for all $q$. Thus,
for a fixed point that is stable for $q=0$, the fixed point remains
stable so long as condition \ref{stab}, or condition $\ref{newq}$ for
$n=2$, is satisfied.  

We also note that, for the case $n=2$ and for $q\leq 1$, a sufficient
condition to conclude that there is only a single fixed point, which
is stable,
is $\Det\, \J>0$ and
$J_{EE}^{2} < J_{IE} J_{II}$, which can be seen as follows.  The
determinant condition is $\Det \left(\bPhiy\J-\1\right)>0$. 
  We note that, for an arbitrary 2-dimensional matrix $\M$, $\Det\,
  \left(\M - \1\right) = \Det\, \M -  \Tr\, \M + 1$.  Thus, the
  determinant condition is $\Det\, \bPhiy\J>\Tr\, \bPhiy\J-1$.  Since
  $\Det\, \bPhiy>0$ (because firing rates and $\alpha$ are $>0$), this
  condition will be satisfied if $\Det\, \J>0$ 
  and $\Tr\, \bPhiy\J<1$.  The trace condition for stability is  
  $\Tr\, \T^{-1}\bPhiy\J<1+q$. But, for $q\leq 1$ and given
  the structures of $\bPhiy$ and $\J$,
  $\Tr\, \T^{-1}\bPhiy\J\leq \Tr\, \bPhiy\J$, so the condition 
  $\Tr\, \bPhiy \J<1$ ensures that the trace condition is also satisfied.
This condition is 
\begin{equation}
J_{EE}y_E^{\frac{n-1}{n}} - J_{II}y_I^{\frac{n-1}{n}}<\frac{1}{n\alpha^{\frac{1}{n}}}
\label{trace}
\end{equation}
For $n=2$, we substitute the solution for $\sqrt{y_I}$ as a function
of $y_E$ (footnote \ref{footcond})
into Eq.~\ref{trace} for $n=2$ to find
$J_{EE}^{2} - J_{IE} J_{II} < \frac{1 + 4 \alpha g_{I} J_{II}}
{4 \alpha y_E}$.  Since the right side is
positive, a sufficient condition for this to be true is 
$J_{EE}^{2} < J_{IE} J_{II}$.  Recall that, if there is more than one
fixed point, some will be unstable at $q=0$, and they must remain
unstable for some region of small but finite $q$.  Since this
condition guarantees that any fixed point is stable, we conclude that
there can only be one fixed point, which is stable, when this
condition holds.

In summary, for $q=\tau_I/\tau_E=0$, the network always flows to a
stable fixed point if $\Det\, \J>0$.  For $q>0$, a fixed point that is
stable at $q=0$ remains stable when Eq.~\ref{stab} or, for $n=2$,
Eq.~\ref{newq} is satisfied.  Note that this condition does not ensure
that the network always flows to a stable fixed point; for nonzero $q$
there may be initial conditions outside the basin of attraction of the
stable fixed point or points. A condition that ensures that any fixed
point is stable for $q\leq 1$, and therefore that there is only one
fixed point, is $\Det\, \J>0$ and $J_{EE}^{2} < J_{IE} J_{II}$.  If
there are no limit cycles (stable or unstable), this 
ensures that the network will flow to the stable fixed point.
Excepting Eqs.~\ref{stab}-\ref{newq}, these conditions involve
feedback inhibition being sufficiently strong:
$J_{IE}J_{EI}>J_{EE}J_{II}$, and $J_{IE}>\frac{J_{EE}^2}{J_{II}}$.

In Fig.~\ref{parameters}, bottom row, we will
illustrate the range of $q$'s yielding stability for various parameter
choices with $n=2$.

\subsection{The case $\left(-\J^{-1}\g\right)_E<0$ and supersaturation}
\label{nege}

We consider Eq.~\ref{requ} for $\r$, but substituting $\psi \J$
for $\W$.  We restrict to the case $\Det\, \J>0$, which ensures
a stable fixed point for at least some range of
$\frac{\tau_I}{\tau_E}>0$. We note that for $\Det\, \J>0$,
$\left(-\J^{-1}\g\right)_E<0$ and $\left(-\J^{-1}\g\right)_I<0$ are
equivalent to $\nege<0$ and $\negi<0$, respectively.  

We shall equate increasing or decreasing $c$ with increasing or
decreasing stimulus contrast.  This is based on the fact that the
contrast of a visual stimulus is monotonically (but nonlinearly)
related to the firing rate of the inputs to V1 from the lateral
geniculate nucleus (LGN) (\eg\ \citeA{Ohzawa_etal85}). 

In simulations, we find that if $\nege<\negi<0$ for $g_E=g_I$, then
$r_E$ grows with $c$ for a range of $c$ considerably beyond the
transition from supralinear to sublinear behavior, but ultimately
peaks and is pushed back to 0 with increasing $c$ (see
Fig.~\ref{parameters}A).  The inputs to cortex have limited dynamic
range (\eg\, stimulus contrast cannot increase beyond 100\%), and so we
imagine that this circuit may model cortex but that the maximal input
strength seen biologically cannot drive excitatory responses too far
beyond their peak.  The decrease in response with increasing contrast
after a peak response is referred to as ``supersaturation'', and is
seen in virtually all V1 cells for contrasts larger than about 75\%
\cite{Ledgeway_etal05,Li_Creutzfeldt84,Tyler_Apkarian85,Peirce07}.
This model behavior provides one possible explanation for
supersaturation, although supersaturation might also in part reflect a
supersaturation of inputs, \eg\ if feedforward
inhibition~\cite{Bruno11} overtakes feedforward excitation with
increasing contrast.

Here we analyze this behavior.  We shall find that (1) if $\r$ is a
stable fixed point, then $\frac{\partial r_E}{\partial c}$ and
$\frac{\partial r_I}{\partial c}$ are
negative precisely when $\nege < -\frac{g_E}{n \psi k^{\frac{1}{n}}
  r_I^{\frac{n-1}{n}}}$ and $\negi < -\frac{g_I}{n \psi
  k^{\frac{1}{n}} r_E^{\frac{n-1}{n}}}$, respectively (and so in
particular can only be negative if $\nege<0$ or $\negi<0$,
respectively); (2) if $\nege<0$, then there is a stable fixed point
with $r_E=0$ at a finite positive value of $c$, which we calculate;
and, (3) for $n=2$, if in addition $\nege < \frac{g_E^2}{g_I^2}\negi$,
then the set of fixed point excitatory rates,  $r_E(c)$, has a maximum
(where $\frac{\partial r_E}{\partial c}=0$ and $\frac{\partial^2
  r_E}{\partial c^2}<0$), which is typically the peak of $r_E$  
 before it is pushed to zero, and we
calculate the corresponding $c$ and peak value of $r_E$.

The condition $\nege<0$ states that the linear term in $c$ in the
high-$c$ expansion for $r_E$ (Eq.~\ref{highc}) is negative, driving
$r_E$ to zero. A related criterion for supersaturation was
noted by \citeA{PersiVreeswijk:2011}, who studied a high-dimensional
ring network with power-law input/output functions much like the ring
model studied here, though with varying connectivity widths for the
four connection types and different power laws for E vs. I cells. They
assumed a stimulus of orientation $\theta$ gave feedforward input $I_X
e^{-\frac{d(\theta,\theta')}{2\sigma_{X,\text{LGN}}^2}}$ to cells of
  population $X$ (E or I) and preferred orientation $\theta'$.  For
  $I_E=I_I$, their criterion for supersaturation was $\Jei
  \sigma_{I,\text{LGN}} - \Jii \sigma_{E,\text{LGN}} >0$, which
  appears closely related to our criterion $\nege<0$, \ie\ $\Jei g_I -
  \Jii g_E >0$.  Our other condition, $\nege < \frac{g_E^2}{g_I^2}\negi$,
  states that the linear term in the expansion for $r_I$ is either
  positive, or not so negative as to disrupt the ability of inhibition
  to drive $r_E$ to zero.

These results suggest, but do not prove, that $r_E$ will be driven to
zero for arbitrary $n$ whenever $\nege<0$ (although there is a stable
fixed point with $r_E=0$ at finite $c$, we have not proven that it is
the \emph{only} stable fixed point).  We note that $r_I$ can never be zero
for finite $c$ if $g_I>0$ (since $r_E\geq 0$), so given $g_I>0$, $r_I$ can
never be driven to zero with increasing $c$ even for $\negi<0$.

For $\nege<\negi<0$, $g_E=g_I$, we find in simulations that $r_I$ only
increases with increasing $c$ (see
Fig.~\ref{parameters}A).  We speculate that, for $\nege<0$ and $\nege
< \frac{g_E^{f(n)}}{g_I^{f(n)}}\negi$, where $f(n)$ may equal $n$ or
may equal 2, $r_E$ can never become large enough to set
$\frac{\partial r_I}{\partial c}<0$, while $r_I$ always becomes large enough to set
$\frac{\partial r_E}{\partial c}<0$ and so ultimately to drive $r_E$ to zero.

When $\negi<\nege<0$ for $g_E=g_I$, we find unbiological behavior in simulations in which both $r_E$ and $r_I$ jump to very high levels at very low $c$,
after which $r_E$ monotonically decreases and is ultimately pushed to 0
(see Fig.~\ref{parameters}E).  Numerical calculations suggest a
discontinuity at the jump, which may explain why our calculations do
not find a zero of $\frac{\partial r_E}{\partial c}$ for real positive $c$ in this
case.  We have not tried to analyze this behavior.

\subsubsection{When can $r_E$ or $r_I$ decrease with contrast?}
\label{sec-supersat}

We define the matrix $\bPhir=n
k^{\frac{1}{n}} \mat{r_E^{\frac{n-1}{n}}}{0}{0}{r_I^{\frac{n-1}{n}}}=\frac{1}{\psi}\bPhiy $.
Then a simple calculation shows that $\frac{\partial \r}{\partial c} =
\bPhir(\psi\J\frac{\partial \r}{\partial c}+\g)$ or  
$\frac{\partial \r}{\partial c}=\left(\1 - \psi\bPhir\J\right)^{-1}\bPhir\g$, which gives
\begin{equation}
\frac{\partial \r}{\partial c}=
\frac{n k^{\frac{1}{n}}
\vectwo{r_E^{\frac{n-1}{n}}\left(\Omega_E n \psi k^{\frac{1}{n}}r_I^{\frac{n-1}{n}}+g_E \right)}
{r_I^{\frac{n-1}{n}}\left(\Omega_I n \psi k^{\frac{1}{n}}r_E^{\frac{n-1}{n}}+g_I \right)}}
{\Det\, \left(\1-\psi\bPhir\J\right)}
=
\frac{\frac{1}{\psi}n \alpha^{\frac{1}{n}}
\vectwo{y_E^{\frac{n-1}{n}}\left(\Omega_E n \alpha^{\frac{1}{n}}y_I^{\frac{n-1}{n}}+g_E \right)}
{y_I^{\frac{n-1}{n}}\left(\Omega_I n \alpha^{\frac{1}{n}}y_E^{\frac{n-1}{n}}+g_I \right)}}
{\Det\, \left(\1-\bPhiy \J\right)}
\end{equation}
Stability requires that 
$\Det\, \left(\1-\bPhir\W\right)>0$. Thus, this expression shows that,
for a stable fixed point, $r_E$
or $r_I$ decrease with contrast precisely when
\begin{eqnarray}\label{maxrE}
\nege &<& -\frac{g_E}{n \psi k^{\frac{1}{n}} 
    r_I^{\frac{n-1}{n}}} = -\frac{g_E}{n \alpha^{\frac{1}{n}} y_I^{\frac{n-1}{n}}} 
\qquad \left(\frac{\partial r_E}{\partial c}<0\right)\\ 
\negi &<& -\frac{g_I}{n \psi k^{\frac{1}{n}} 
    r_E^{\frac{n-1}{n}}} = -\frac{g_I}{n \alpha^{\frac{1}{n}} y_E^{\frac{n-1}{n}}} 
\qquad \left(\frac{\partial r_I}{\partial c}<0\right)
\end{eqnarray}

\subsubsection{The $c$ at which $r_E$ becomes 0}

\label{rezerosec}
The $c>0$ at which $r_E$ first becomes 0 with increasing $c$ can be
determined as follows. 
First, at this $c$, $r_I=c g_E/\psi J_{EI}$, because this is the
value of $r_I$ that sets the input to $r_E$ to zero when $r_E=0$. The
equation for the $r_I$ steady state then yields $\frac{c g_E}{\psi
  J_{EI}} = k\left(c g_I- c g_E\frac{J_{II}}{J_{EI}}\right)^{n} = k
c^n \left( \frac{-\nege}{J_{EI}}\right)^{n}$.  The right side gives
zero unless $\nege<0$, so a solution for $c\neq 0$ exists only for
$\nege<0$.  In this case, one can solve to find $c = J_{EI}
\left( \frac{g_E}{k\psi \left(-\nege\right)^n} \right)^{\frac{1}{n-1}}
$. This corresponds to $\alpha=
\frac{J_{EI}^{n-1}g_E}
{\left(-\nege\right)^n}$ or 
$\beta=\frac{- \nege}{\left(J_{EI}^{n-1}g_E\right)^{\frac{1}{n}}}$.\footnote{
Once $r_E$ has been pushed to zero, for increasing $c$, $r_I$
continues to increase according to $r_I=k(c g_I-\psi J_{II}r_I)^n$,
which for $n=2$ has the solution $r_I=\frac{\left(\sqrt{1+4 c g_I \psi
J_{II} k^2}-1\right)^2}{4 k \psi^2 J_{II}^2}$, and $r_E$ remains 0.}

Note that any fixed point $y_E=0$, $y_I>0$ is stable
for any $q$ since the Jacobian matrix is ${\cal
  J}=n\alpha^{\frac{1}{n}} \mat{-1}{0}
 {\frac{1}{q}y_I^{\frac{n-1}{n}}\psi J_{IE}} 
{-\frac{1}{q}\left(y_I^{\frac{n-1}{n}}\psi J_{II}+1\right)}$,
  which has two negative eigenvalues (equal to the two diagonal
  entries of ${\cal J}$). 

  This shows that $r_E=0$, $r_I=c g_E/\psi J_{EI}$ is a stable fixed
  point for this value of $c$, but does not rule out the existence of
  other fixed points.

\subsubsection{Peak excitatory firing rate and corresponding contrast}
 \label{sec-cmax}
 
 For the case $n=2$ and $\Omega_E<0$, we can calculate explicitly the
 contrast at which the steady-state excitatory firing rate reaches a
 local maximum, $\frac{\partial r_E}{\partial c} =0$, and the
 corresponding excitatory firing rate. We will refer to this as the
 peak excitatory firing rate.  We are imagining that there is a
 continuous curve of stable fixed points vs.\ contrast stretching from
 zero firing rate for $c=0$ to the fixed point with $r_E=0$ at
 positive $c$ found in the previous section and that the dynamics
 converge to these fixed points, in which case we are finding the
 peak steady-state excitatory firing rate.  This has been the case in all the
 simulations we have studied.  However, for other parameters it is
 possible that other stable fixed points may appear, that the maximum
 occurs at negative $c$ and thus is biologically irrelevant (further
 discussed below), and/or that the 
 local maximum found here is an unstable fixed point.

To find the peak excitatory firing rate and the
$c$ at which it occurs for the case $n=2$, we have
to solve the steady-state equations $k(\psi \J \r + c \g)^{.2} = \r$,
together with the extremum condition $\frac{\partial r_E}{\partial c}
=0$, for the three variables $r_E$, $r_I$ and $c$. The extremum
condition is given by \req{maxrE} with equality, which for $n=2$,
yields $\sqrt{4 k \psi^2 r_I} = -\frac{g_E}{\nege}$ (as noted above,
this can only happen when $\nege<0$, which we assume here). In order
to simplify the equations, we will change variables according to  $\r
= \frac{\vc{x}^{.2}}{4 k \psi^2}$, where without loss of generality we
require $\vc{x}$ to be real and positive, ensuring the positivity of
$\r$. We also rewrite the steady-state equations by taking their
square roots as 
\be\label{n2fp}
\J \vc{x}^{.2} - 2 \vc{x} = -4 k \psi c\, \g,
\ee
while the extremum condition now becomes $x_I = -\frac{g_E}{\nege} =  \frac{g_E}{|\nege|}$.
Multiplying \req{n2fp} by $\J^{-1}$, we obtain $\vc{x}^{.2} - 2
\J^{-1} \vc{x} =4 k \psi c \frac{\vct{\Omega}}{\Det\, \J}$ where the
vector $\vct{\Omega}=\vectwo{\Omega_E}{\Omega_I}$. We then multiply the $E$ and $I$ components of the latter equation by $\negi$ and $\nege$, respectively (we assume $\negi\neq 0$), and subtract the results. After simplifying, this yields $\negi \xx_E^2 -\nege \xx_I^2 + 2g_I \xx_E - 2 g_E \xx_I = 0$.  Finally, substituting for $x_I$ using the extremum condition, we obtain $\negi \xx_E^2 + 2g_I \xx_E - \frac{g_E^2}{|\nege|} = 0$, which has one positive solution
\begin{equation}
\xx_E= 
\frac{g_I}{\negi} \lbr \sqrt{1 + \frac{g_E^2}{g_I^2}\frac{ \negi}{ |\nege| } } - 1 \rbr,
\label{kappaeq}
\end{equation}
yielding
\begin{equation}
r_E^{\max} = \frac{\xx_E^2}{4 k \psi^{2}}
\label{remax}
\end{equation}
for the maximum firing rate.
Note that for this to be positive, $\xx_E$ must be real, which means an additional condition needs to be
satisfied, namely $g_E^2 \negi > -g_I^2|\nege|$.  

The contrast, $c^{\max}$, at which $r_E$ peaks can then be solved for using either component of \req{n2fp}, yielding
\bea
c^{\max} \areq \frac{1}{4 k \psi g_{E}}
\left(J_{EI} \frac{g_{E}^{2} }{\nege^{2}} + 2 {\xx_E}  - J_{EE}
  \xx_E^2  \right)       \label{cmax}
  \\
  \areq  \frac{1}{4 k \psi g_I} \lpr J_{EI} \frac{g_E g_I}{\nege^2} + \frac{g_E }{ |\nege|} - J_{IE} x_E^2 \rpr
\label{cmax2}
\eea
with $\xx_E$ given by \req{kappaeq}.
Explicit calculation shows that this is
always a maximum: the $2^{\msub{nd}}$ derivative $\frac{d^2
  r_E}{dc^2}<0$.
$c^{\max}$ is not guaranteed to be positive
-- this is governed by rather complicated conditions on the $g$'s and
$J$'s -- but in practice we have found it to be positive for the
simulation parameters we have used.\footnote{$c^{\max}$ and $\r^{\max}$ correspond to $\alpha= \frac{1}{4 g_{E}}
\left( J_{EI} \frac{g_{E}^{2} 
    }{\nege^{2}} - J_{EE} \xx_E^2 + 2 {\xx_E} \right)$ and
$y_E=\frac{\psi}{c^{\max}}r_E^{\max}=\frac{\xx_E^2 g_E}{\left(
    J_{EI}\frac{g_{E}^{2} }{\nege^{2}} - J_{EE} \xx_E^2 + 2
   {\xx_E} \right)}$.  However, note that this 
is not a maximum of the $y_E$ vs.\ $\alpha$ curve, but rather occurs
for $\alpha$ higher than that peak, where the curve has a negative
slope.  We saw in section~\ref{normsec} 
that $\frac{\partial \r}{\partial c}=\frac{(n-1)\alpha}{\psi}\left(
  \frac{d\y}{d\alpha} + \frac{\y}{(n-1)\alpha}\right)$, so 
$\frac{\partial \r}{\partial c}=0$ implies $\frac{d\y}{d\alpha} =
-\frac{\y}{(n-1)\alpha}$, \ie\ $\y$ is locally evolving as $\alpha^{-\frac{1}{n-1}}$.
}

When $c^{\max}$ is positive and $r_E^{\max}$ is indeed the peak steady-state
excitatory firing rate, then
Eqs.~\ref{remax} and \ref{cmax2}
show that both the maximum excitatory firing rate that
can be achieved by the network and the contrast at which this maximum
is achieved decrease with increasing $\psi$.  In this case, when the 2-D reduced
model, Eqs.~\ref{twode}-\ref{twodi}, accurately captures the
high-dimensional model, Eqs.~\ref{highde}-\ref{highdi}, then in the
high-D model, if the stimulus is
widened or a second stimulus is added, the maximum excitatory firing
rate will decrease and will occur at a lower contrast.

In sum, for $\Det\,\J>0$ and $n=2$, the steady state solution for
$r_E$ has a maximum value as a function of $c$ (\ie, a point with
$\frac{\partial r_E}{\partial c} =0$ and $\frac{\partial^2 r_E}{\partial c^2} <0$), given by
Eq.~\ref{remax}, precisely when (1) $\nege<0$ and (2) $\frac
{g_I^2}{g_E^2} \nege < \negi$.

\subsection{Steady-state solutions for different parameter regimes}
\label{solutions}

\begin{figure}[htp]
\centering
\vspace {-1 cm}
\hspace{-0.5 cm}
\includegraphics[width=16.3cm] {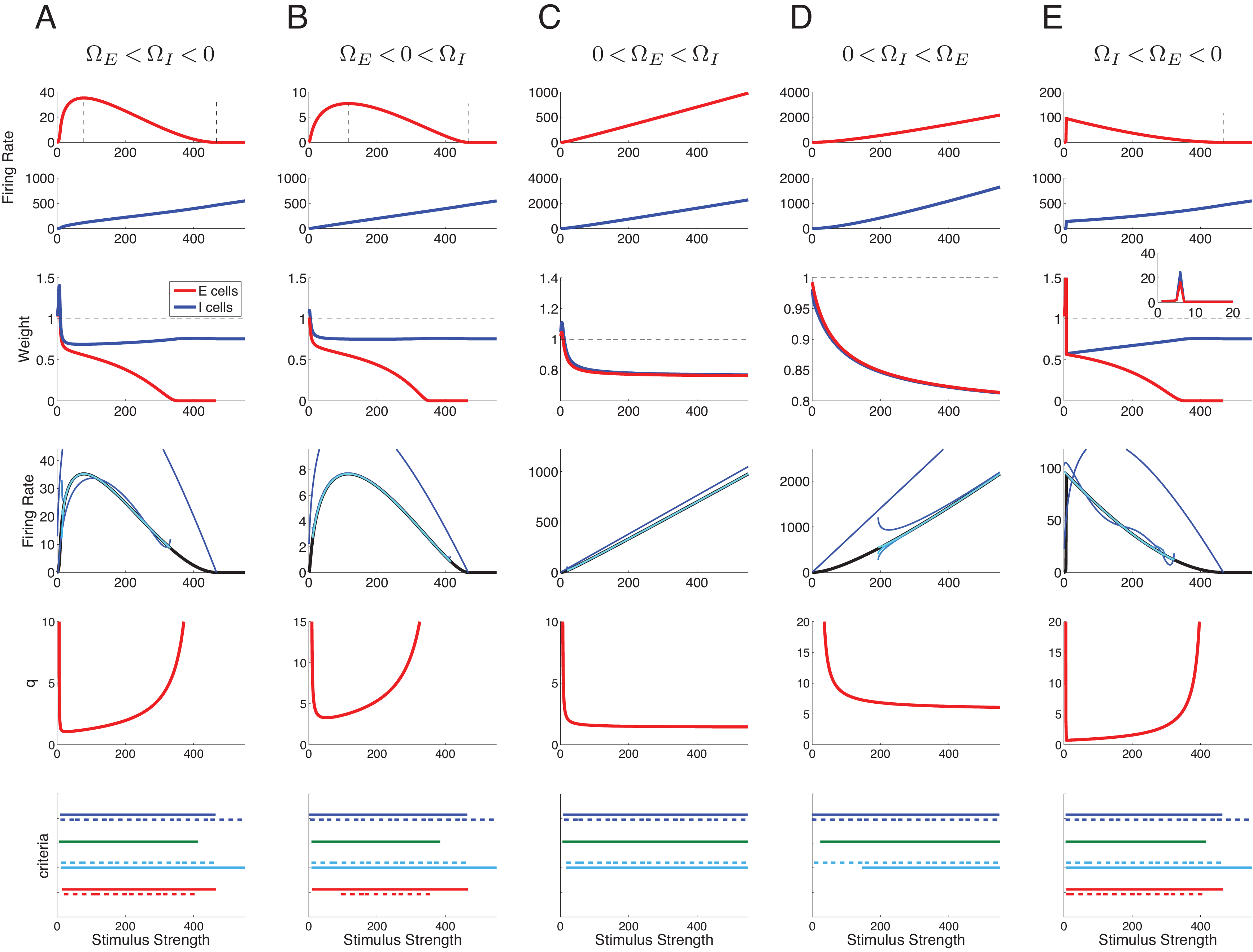}
\vspace{ 0 cm}
\caption[Behavior of the 2D Model in Different Parameter Regimes] {\small
  {\bf Behavior of the 2D Model in Different Parameter Regimes}. 
  Each column corresponds to a different connectivity matrix $\J$,
  corresponding to different conditions on $\Omega_E$ and $\Omega_I$
  as indicated at top.  In all cases, $\Det\, \J>0$, $n=2.0$, $k=0.04$, and 
  $g_E=g_I=1$. 
The first column uses the same parameters as the 2-D reduced model in
Fig.~\ref{analyticRing}. In all figures the horizontal axis is
stimulus strength $c$; Fig.~\ref {parameters2d-lowc} shows all 
plots (and one additional set of plots) for the
smaller range $c \in [0,40]$. 
\thisfloatpagestyle{empty}
{\bf Top row}: $E$ (red, top) and $I$ (blue, bottom) firing rates,
$r_E$ and $r_I$, at 
fixed point.  For cases with $\Omega_E<0$,
dashed vertical lines indicate analytic calculations for $c$ at which
$r_E$ goes to zero (Sec.~\ref{rezerosec}) and, for $\Omega_E<\Omega_I$,
at which $r_E$ peaks (Eq.~\ref{cmax2}).
{\bf Second Row}: Weights reflecting supralinear (weight
$>1$) or sublinear (weight $<1$) summation in an equivalent ring
model, computed as in Fig.~\ref{analyticRing}B.  Red and blue
indicate $E$- and $I$-subnetworks, respectively. Inset in column E
shows supralinear responses at low values of $c$.
{\bf Third Row}: Iterative solutions for $r_E$ in the high-contrast
regime (Eq.~\ref{hione}). We plot
$r_E[t]=y_E[t]c/(\psi \|\J\|)$ vs.\ $c$, for $t =$ 1, 5, 10, 14, 19 iterations (blue to cyan curves); black curves are exact solutions.  Iterative
solutions are shown only over the range for which they are
real. (Iterative solutions in the low-contrast regime are shown in Fig.~\ref {parameters2d-lowc}.)
{\bf Fourth Row}: Values of $q=\tau_I/\tau_E$ separating regions in
which fixed point is stable (below red line) vs.\ unstable (above red
line).
{\bf Fifth Row}: Horizontal lines showing the extent of the sublinear
regime according to the different definitions introduced in Sec.~\ref
{sec:normcriter}. Blue and red lines ($E$ component solid, $I$
component dashed): definitions 1
(normalization in corresponding high-dimensional ring model) and 5 ($r$
a sublinear function of $c$), respectively. Green line: definition 2
(excitatory subnetwork unstable). The cyan lines show the range where
the modulus of each eigenvalue of the Jacobian is $>1$; sublinear
regime according 
to definition 3 (instability of low-contrast iterative solution) or 4
(stability of high-contrast iterative solution) is the region in which
either (def.~3) or both (def.~4) lines are present. 
Parameters used: $\psi=0.774$ or, for two-grating case in 2nd row, $\psi=1.024$
 (the values of $\Psi$
in Fig.~\ref{analyticRing});
$\Jei=1.3$;  
$\Jee=2.5$, except $0.8$  in (D);
$\Jii = 1.0$, except $2.2$ in (C) and $5.0$ in (D);
$\Jie = 2.4, 4.7, 4.7, 3.6, 2.2$ in (A) to (E), respectively. 
\smallskip}
\label{parameters}
\end{figure}

\begin{figure}[htp]
\centering
\vspace{-0.5 cm}
\hspace{-0.5 cm}
\includegraphics [width=16.3cm] {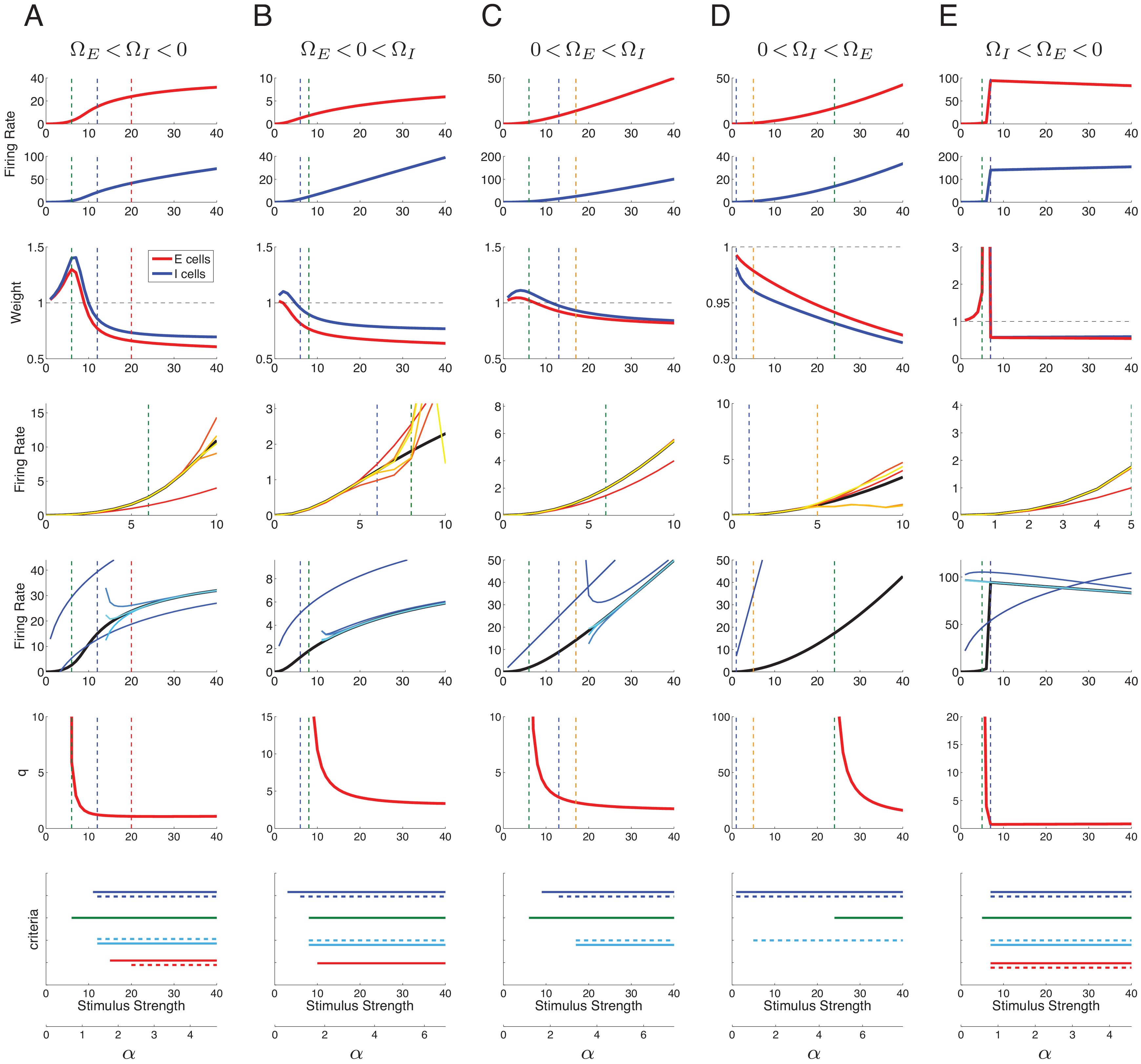}
\vspace{ 0 cm}
\caption[Crossover of the 2D Model to the High Contrast Sublinear Regime in Different Parameter Regimes]{\small
  {\bf Crossover of the 2D Model to the High Contrast Sublinear Regime for Different Network Parameters}. 
The plots in this figure are the same as the ones in
Fig.~\ref{parameters}, except that (1) only the range of stimulus
strengths, $c$,   from 0 to 40 is shown, to highlight the transition
to the sublinear regime as $c$ grows and (2) we also illustrate the
low-contrast iterative solutions for $r_E$, which have been inserted
as the third row (conventions as for high-contrast iterative
solutions, except here red to yellow curves represent 1 to 19 iterations). See the caption of  Fig.~\ref{parameters} for explanation of plots and parameters. The extra horizontal axes at the bottom translate the stimulus strengths into values of $\alpha$ as defined in \req {dimensionless}. 
In addition, vertical dashed lines in the first to fifth rows indicate
the transition points to the sublinear regime, according to the
different definitions introduced in Sec.~\ref {sec:normcriter} and
illustrated in the bottom row, with definitions 1 to 5 corresponding to colors blue,  green, orange, cyan  and red, respectively (for definitions 1 and 5 the line is drawn at the point where the condition holds for both $E$ and $I$ components). 
The values of the $\alpha$'s at these transition lines, in the order mentioned,  are $(1.4,0.7,1.4,1.4,2.4)$, $(1.0,1.4,1.4,1.4,16.3)$, $(2.4,1.1,3.1,3.1,-)$, $(0.2,4.7,1.0,28.5,-)$, and $(0.8,0.6,0.8,0.8,0.8)$ in columns A to E, respectively.
Notice that the transitions to the sublinear regime typically happen
for $\alpha \sim 1$, as expected. 
\smallskip
}
\label{parameters2d-lowc}
\end{figure}

In Figs.~\ref{parameters}-\ref {parameters2d-lowc} we illustrate model
behavior, as a function of stimulus strength $c$, for 5 parameter
regimes, with $\Det\, \J>0$, $n=2$, and $g_E=g_I$ in all cases.  In
Fig.~\ref{parameters} we illustrate behavior across a large range of
$c$, sufficient to see overall model behavior.  To better illustrate
the region around the transition to normalizing behavior, in Fig.~\ref
{parameters2d-lowc} we replot Fig.~\ref{parameters} but restricting to
the range $c=0$ to $40$.  The 5 illustrated parameter regimes are:
$\Omega_E<0$ and $\Omega_I<0$, with either $\Omega_E<\Omega_I$
(Figs.~\ref{parameters}-\ref {parameters2d-lowc}A) or
$\Omega_E>\Omega_I$ (Figs.~\ref{parameters}-\ref
{parameters2d-lowc}E); $\Omega_E<0$ and $\Omega_I>0$
(Figs.~\ref{parameters}-\ref {parameters2d-lowc}B); and $\Omega_E>0$
and $\Omega_I>0$, with either $\Omega_E<\Omega_I$
(Figs.~\ref{parameters}-\ref {parameters2d-lowc}C) or
$\Omega_E>\Omega_I$ (Figs.~\ref{parameters}-\ref
{parameters2d-lowc}D).  We chose parameters relatively arbitrarily, by
starting with a set of parameters that had worked well in simulations
of the ring model (Fig.~\ref{analyticRing} and
Fig.~\ref{parameters}-\ref {parameters2d-lowc}A) and changing small
sets of parameters to change the regime.  However in small amounts of
studies of other parameters in the different regimes we have found
behaviors to be similar to those illustrated, with one exception.  For
$\Omega_E<0$, the transition to sublinear scaling can occur and the
excitatory firing rate can peak and be driven to zero without the
excitatory rate ever reaching a level at which the excitatory
subnetwork is unstable.  This would be manifested in the figures as
stability for all possible values of $q$ (fifth row, described below).
In simulations we only encountered this for $\Omega_I>0$ with
relatively weak $\Jee$ (note that $\Omega_I>0$ includes the case
$\Jee=0$, and presumably sufficiently small $\Jee$ behaves similarly
to that case), but it can occur for $\Omega_I<0$ as well.  The
conditions in which the excitatory subnetwork does not become unstable
are discussed more generally in Section \ref{sec:normcriter}.

For each set of parameters, we first illustrate firing rates (top
row), with red and blue indicating $r_E$ and $r_I$ respectively.
As expected, parameters with $\Omega_E<0$ (columns A,B,E) all show
$r_E$ eventually pushed to zero with increasing $c$, while those with
$\Omega_E>0$ (columns C,D) show $r_E$ moving toward linear growth with
increasing $c$.  The combination $\Omega_I<\Omega_E<0$ (column E)
leads, as mentioned previously, to unbiological behavior in which both
E and I rates abruptly jump (discontinuously, in numerical
calculations with $c$ discretized in 0.00001 steps) to high rates at low
$c$, after which $r_E$ monotonically falls with increasing $c$.

If biology is represented by a case with $\Omega_E<0$ and
$\Omega_E<\Omega_I$ (columns A,B), we imagine the dynamic range of
cortex, corresponding to the dynamic range of the firing rates of the
inputs to cortex, represents a smaller range extending up to and slightly
beyond the point at which $r_E$ peaks as a function of $c$, as
discussed in Section~\ref {nege}.  
An example is the range through $c=100$ in Fig.~\ref{analyticRing}A,
reduced model, 1 stimulus, which uses essentially the same parameters
as Figs.~\ref{parameters}-\ref
{parameters2d-lowc}A.  Biologically, supersaturation begins at
high contrasts, \eg\ 75\% \cite{Li_Creutzfeldt84}, well beyond the
contrasts (10\%-20\%) at which the transition from sublinear to supralinear
summation \cite{Heuer_Britten02,Ohshiro_etal11} or from surround
facilitation to surround suppression
\cite{Sengpiel_etal97,Polat_etal98} occur.  That is, while
the dynamic range of cortex ends shortly after supersaturation is
seen, much of this dynamic range exhibits normalizing behavior.
Similarly, the model shows a broad dynamic range between the onset of
normalization and of supersaturation for most parameter choices we have
explored, the only exception again being the case $\Omega_E<0$ and
$\Omega_I>0$ for small $J_{EE}$.

We next illustrate normalization weights (second row), computed just
as in Fig.~\ref{analyticRing}B, right column, so that weights $>1$ (weights $<1$) indicate
supralinear (sublinear) summation of responses to two orthogonal gratings of equal strength in the corresponding ring model.  All but the case $\Omega_E>\Omega_I>0$
show a regime of supralinear summation for very low contrasts
(behavior in all cases is sublinear for $c>10$), although the
supralinear behavior is weak for $\Omega_I>0$.

The third and fourth rows of Fig.~\ref
{parameters2d-lowc} illustrate the 
iterative solutions that stem from the scaling solutions in the low-
and high-contrast regimes (the high-contrast iterative solutions are
also illustrated in the third row of Fig.~\ref{parameters}).
The values of $\J$ used (listed in legend
of Fig.~\ref{parameters}) are not normalized to have $\|\J\|=1$, so
for these iterations we take $\hat \J=\J/\|\J\|$ 
where $\|\J\|$ is the 2-norm of $\J$ (the maximum singular
value of $\J$), and redefine $\alpha$ and $\vy$ such that $\alpha=k c^{n-1}\psi \|\J\|$ and $\r=\frac{\vy c}{\psi \|\J\|}$.
We show the iteration results
as $r_E$ vs.\ $c$. The reason for this rescaling is that, as discussed in Section~\ref{sec:scaling}, with this definition of $\alpha$ the transitions to the sublinearly normalizing regime happen at $\alpha \sim 1$, irrespective of $\|\J\|$. 

The small-$\alpha$ (low contrast) iterations are shown in the third
row of Fig.~\ref{parameters2d-lowc}.  Here, we treat the equation
$\y=\alpha(\hat \J\y+\g)^{.n}$ as the recurrence relation
\begin{equation}
\y[t]=\alpha(\hat \J\y[t-1]+\g)^{.n}
\label{lowone}
\end{equation}
Iteration of this recurrence relation generates higher- and
higher-order approximations to Eq.~\ref{yalpha} for the steady state.
We use
the starting condition $y[0]=0$.  Results are shown for numbers
of iterations ranging from 1 to 19 (red through yellow colors).  As
few as 5 iterations gives a 
good approximation for small $c$, while increasing the number of
iterations to 19 adds little.  The low-contrast iterations all fail
before or very slightly after $c=10$, which corresponds to $\alpha$ in the
range 1.4 to 2.4 across the parameters. That is, the failure occurs
for $\alpha \sim 1$, as expected.

For the high-$\alpha$ or small-$\beta$ (high contrast) case, we treat
the equation $\y= \hat\J^{-1}\left(-\g+\beta\y^{.\frac{1}{n}}\right)$ as
the recurrence relation:
\begin{equation}
\y[t]= \hat\J^{-1}\left(-\g+\beta\y[t-1]^{.\frac{1}{n}}\right)
\label{hione}
\end{equation}
Iterations generate approximations to Eq.~\ref{highalpha} for the
steady state.
We use as starting conditions $y_E[0]=0$ with $y_I[0]=0$ for
$\Omega_E>0$, $y_I[0]=g_I/ \hat J_{EI}$ for $\Omega_E<0$.  For $\Omega_E<0$,
using $\y[0]=0$ would give complex solutions.  We instead use as a
starting condition the value of $\y$ when $y_E$ reaches zero with
increasing $c$.  Recall that $\beta$
increases with decreasing $c$, $\beta=1/\sqrt{\alpha}$.
For small
$\beta$ (large contrasts) we expect a basin of attraction about the
fixed point, such that the iterations will converge to the fixed point
for initial conditions in the basin of attraction.  As $\beta$ increases
(contrast decreases) the basin of attraction should disappear for
$\beta \sim 1$, so that convergence will fail for any initial condition.

The third row of
Fig.~\ref{parameters} (fourth row of Fig.~\ref{parameters2d-lowc}) illustrates
these high-contrast solutions, with blue through cyan colors
corresponding to 1 through 19 iterations.
Again, 5 iterations do about as well as
larger numbers of iterations.  The iterations give good approximations
for high $c$ but, for $\Omega_E<0$, fail for larger $c$ as $r_E$
approaches zero. $\beta$ is very small for these large $c$'s, so this
presumably represents the initial conditions no longer being in the
basin of attraction of the fixed point. For low $c$ failure of
convergence is expected for $\beta \sim 1$, although
problems with the basin of attraction could also arise.  None of the
iterations work for $c$ below about 9 or 10, corresponding to $\beta$
roughly above .65 to .85, with the exception of column E. In
that column, the largest number of iterations works down to the jump
in $r_E$ and $r_I$, which occurs at about $c=5.435$ for the given
parameters, or $\beta$ around 1.1.  In column D the iterations do not
work below $c$ about 190, which corresponds to $\beta$ above about 0.15, a
somewhat lower value than expected for the $\beta$ at which iterations
fail.

In the fourth row of Fig.~\ref{parameters} (fifth row of Fig.~\ref{parameters2d-lowc}), we show the value
of $q=\frac{\tau_I}{\tau_E}$ that divides stability (values below curves)
from instability (values above curves) of the fixed point, according
to Eq.~\ref{newq}.  In all  
cases except $\Omega_I<\Omega_E<0$, the fixed point remains stable for
$q<1$ across the range of studied stimulus strengths, indicating that
fine tuning or unreasonably small values of $q$ are not required.

Finally, the fifth row of Fig.~\ref{parameters} (sixth row of
Fig.~\ref{parameters2d-lowc}) shows the extent of
the sublinearly normalizing regime. Specifically, the solid and dashed
blue horizontal lines indicate the range of  stimulus strengths for
which $\frac{\partial r_E}{\partial \psi}<0$ and $\frac{\partial
  r_I}{\partial \psi}<0$, respectively. As discussed in
Section~\ref{normsec}, these conditions are roughly equivalent (so
long as the approximate ansatz for 2-D reduction introduced in
Section~\ref{sec:reduction} is valid) to sublinear normalization of $E$
and $I$ subnetworks in the full high-dimensional ring network
considered in Section~\ref{sec:reduction}.\footnote{Note that the onset
  of the conditions $\frac{\partial r_X}{\partial \psi}<0$ 
(vertical dashed blue lines in 2nd row,
Fig.~\ref{parameters2d-lowc}, corresponding to onset of blue lines in
bottom row)
occurs for slightly higher $c$ than the onset of normalizing behavior
(2nd row, Fig.~\ref{parameters2d-lowc}). This is because in the 2nd
row we are 
assaying normalization in response to a finite-strength (equal-contrast)
2nd stimulus, for which, as discussed in Section \ref{normsec}, the
condition for normalization becomes $\int_{\Psi_{\msub{init}}}^{\Psi_{\msub{final}}}
d\Psi \frac{\partial r_X}{\partial \Psi} <0$.}
The other horizontal lines in
the plots of this row show the extent of the sublinear regime
according to other criteria introduced in the next subsection, and are
explained there.

\begin{figure}[htp]
\centering
\vspace{-0.5 cm}
\hspace{-0.5 cm}
\includegraphics[width=16.3cm] {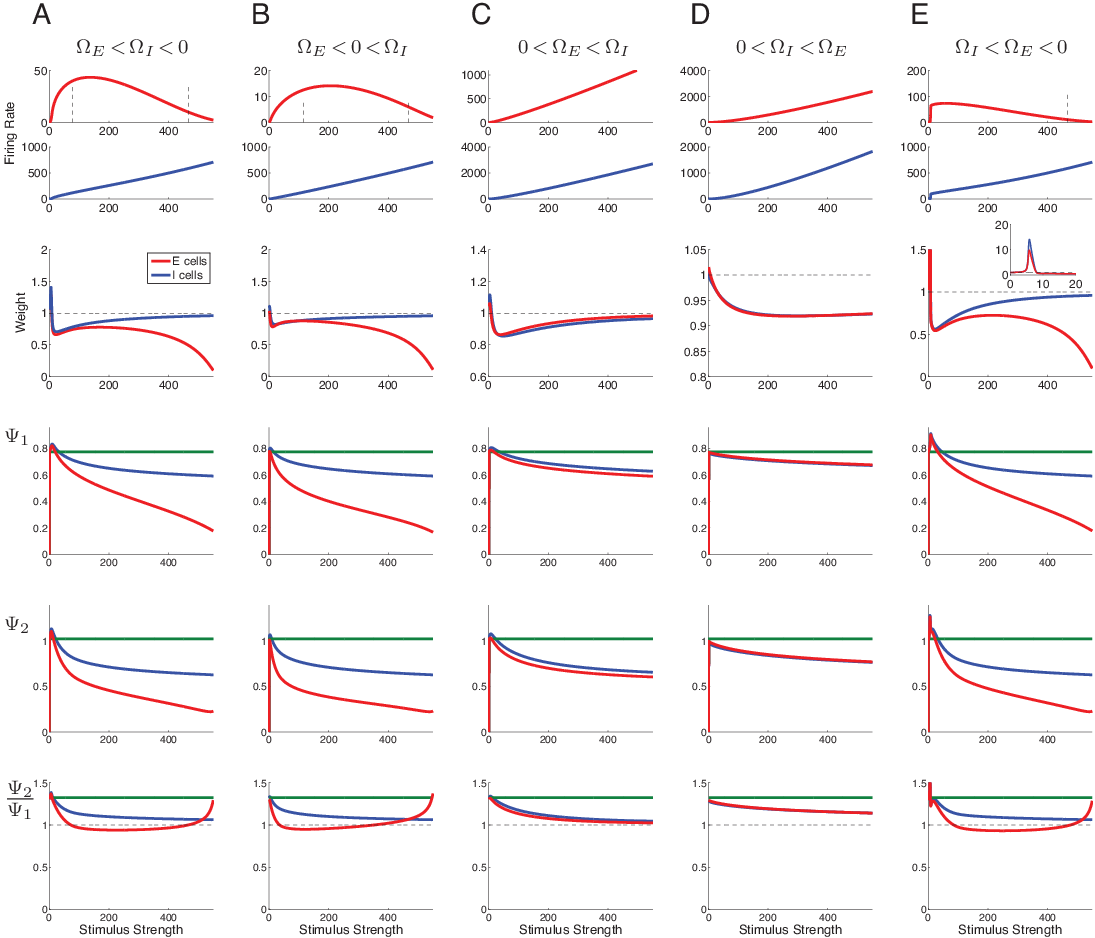}
\vspace{ 0 cm}
\caption[Behavior of the Full Ring Model in Different Parameter Regimes] {\small
  {\bf Behavior of the Full Ring Model in Different Parameter Regimes}. 
Behavior of the steady state of the ring  network of Sec.~\ref{sec:reduction}, in the same parameter regimes as in Fig.~\ref{parameters}. The ring network's connectivity matrix is given by \req{wequ}, with different $\J$'s in different columns equal to those in the corresponding column of Fig.~\ref{parameters}. The rest of the parameters are the same as in the left column of Fig.~\ref{analyticRing} (in particular, all parameters of column A match those of Fig.~\ref{analyticRing}, left column).
  The signs and orderings of $\Omega_E$ and $\Omega_I$ are indicated on the top of each column.  
 In all figures the horizontal axis is
stimulus strength $c$.
\thisfloatpagestyle{empty}
{\bf Top row}: $E$ (red) and $I$ (blue) firing rates, $r_E(\theta =0)$ and $r_I(\theta = 0)$, at
fixed point.  For cases with $\Omega_E<0$,
dashed vertical lines indicate analytic calculations for $c$ in the
2-D reduced model at which $r_E$ goes to zero (Section \ref{rezerosec})
and, for $\Omega_E<\Omega_I$, at which $r_E$ peaks (Eq.~\ref{cmax2}).
{\bf Second Row}: Weights reflecting supralinear summation (weight
$>1$) or sublinear summation (weight $<1$)  computed as in Fig.~\ref{analyticRing}B.  Again, red and blue
indicate $E$- and $I$-subnetworks, respectively. Inset in column E
shows supralinear responses at low values of $c$.
{\bf Third Row}:
The red and blue curves show $\Psi_E \equiv  {\vec{w} \cdot
  \hat{\r}_{E}}$ and $\Psi_I \equiv  {\vec{w} \cdot  \hat{\r}_{I}}$,
which we approximated by $\psi = {\vec{w} \cdot  \hat{\g}^{.n}}$
(green lines) in the 2-D reduction for the case of a one-grating stimulus (see the discussion at the end of Sec.~\ref{solutions}).  
{\bf Fourth Row}: 
The same as the third row, but for two-grating stimuli. 
{\bf Fifth Row}: 
The red, blue and green curves show the ratios of the red, blue and green curves in the fourth row (two gratings) to those in the third row (one grating), respectively. 
\smallskip}
\label{parameters-Nd}
\end{figure}

In Fig.~\ref{parameters-Nd} we examine
the quality of the approximate 2-dimensional
reduction of the high-dimensional ring model (Section~\ref{sec:reduction})
in the five different parameter regimes of
Fig.~\ref{parameters}. In the top two rows, we have plotted the peak
responses of the 
$E$ and $I$ subnetworks and their respective normalization weights for
the original high-dimensional ring network of
\reqs{ringE}--(\ref{ringI}) with the connectivity given by
\req{wequ}, which can be compared with those in the top
two rows of Fig.~\ref{parameters}.  We see that the 2-D model captures the
behavior of the original model very well.  
Recall that the approximation involved in the reduction to the 2-D
model involved taking the ratios $\frac{\vec{w}_{XY} \cdot
  \hat{\r}_{Y}}{\vec{w}_{EE} \cdot  \hat{\r}_{E}}$ in \req{jxydef} as
constants, independent of parameters of the input stimulus (\eg\ its shape and strength), and absorbing all such dependencies into
$\Psi = {\vec{w}_{EE} \cdot  \hat{\vr}_{E}}$. More generally, we could
have defined $\Psi_{XY} = {\vec{w}_{XY} \cdot  \hat{\r}_{Y}}$ (with
$X,Y \in \{ E, I\}$).  In the model of \req{wequ}, the vectors
$\vec{w}_{XY}$ are independent of $X$ and $Y$ by construction: 
$\vec{w}_{XY}\equiv \vec{w}$ where $\vec{w}$ has elements
$\vec{w}(\theta) \equiv \exp\!\lpr 
{-\frac{d(0,\theta)^{2}}{2\sigma_{\msub{ori}}^{2}}} \rpr \Delta
\theta$. Therefore
in this case we only have two independent $\Psi_{XY}$, which we relabel
as $\Psi_E \equiv  {\vec{w} \cdot  \hat{\r}_{E}}$ and $\Psi_I \equiv
{\vec{w} \cdot  \hat{\r}_{I}}$. Finally, we also made the
approximation that the response shape curves, $\hat{\r}_X$, can be
well approximated by the shape of the input, $\g$, raised to the power
$n$, and we used the latter to calculate the $\psi$ used in the 2-D
model, \ie\ we took $\psi = {\vec{w} \cdot  \hat{\g}^{.n}}$. The third
and fourth rows of  Fig.~\ref{parameters-Nd} show plots of $\Psi_E$
(red), $\Psi_I$ (blue), and $\psi$ (green) as a function of stimulus
strength, for the case of one or two grating stimuli,
respectively. The bottom row of  Fig.~\ref{parameters-Nd} plots the
ratio of the curves in the fourth row (two gratings) to the
corresponding ones in the third row (one grating). In the discussion
of Section~\ref{normsec} we assumed that $\Psi_{X}$ should typically be
larger for the case of two gratings; we then concluded that sublinear
normalization weights for the $E$ ($I$) subnetwork are hence
roughly equivalent to $r_E$ ($r_I$) being a decreasing function of $\psi$ in
the 2-D reduced model. We see from the bottom row of
Fig.~\ref{parameters-Nd} that in some parameter regimes and for high
enough $c$ this assumption can weakly fail for $\Psi_E$. Surprisingly,
sometimes this failure is accompanied by the condition $\frac{\partial
  r_E}{\partial \psi} <0$ in the 2-D model, and yet the normalization
weights in the full model are sublinear ($<1$). This is due to
  the failure of the other assumption in the ansatz, \ie\ it is due to the fact that
  $\Psi_E \neq \Psi_I \neq \psi$.

 \subsection{Different criteria for crossover to the sublinearly normalizing regime}
\label{sec:normcriter}

As we saw in Section~\ref{normsec}, the condition that the $E$ and $I$ responses in the high-dimensional network be normalizing is roughly equivalent to $\frac{\partial r_E }{\partial \psi} < 0$ and $\frac{\partial r_I }{\partial \psi} < 0$ in the 2-D model, respectively. 
Here, $r_E$ and $r_I$ refer to their values at a stable fixed point.
More generally, we have seen that for sufficiently low stimulus strengths the network is supralinear (with normalization weights $>1$), but switches to a sublinear regime (with normalization weights $<1$) as stimulus strength becomes sufficiently large. 
One can, however, come up with different notions or criteria for the
transition from the supralinear to the sublinear regime as stimulus
contrast grows. Furthermore, since this is typically a smooth
crossover, and not a sharp phase transition, such different criteria
in general do not yield the exact same value of contrast at the transition, although as we will see, they all yield the same order of magnitude for the transition contrast. 
As this crossover is at the heart of the present study, in this
section we set  out  to examine more closely the different criteria
for the transition to the sublinear regime and their
inter-relationships.

We note that the criteria we will examine all involve addition of
stimuli without change in the values of $g_E$ or $g_I$ (or, for
normalization in the high-dimensional ring model, addition of a second
stimulus at a different location on the ring but with the same
relative strengths of inputs to E vs. I cells as the first
stimulus). One can imagine a different kind of sublinearity due to
rectification, in which a stimulus 1 has a large enough
$\frac{g_E}{g_I}$ that by itself it produces $r_E>0$ and $r_I>0$,
while a smaller stimulus 2 that has small $\frac{g_E}{g_I}$ by itself
produces $r_E=0$ and $r_I>0$. Then addition of stimulus 2 to stimulus
1 would reduce $r_E$ and, if the network is in the ISN regime in
response to stimulus 1, reduce $r_I$ \cite{Ozeki_etal09,Tsodyks_etal97},
relative to the response to stimulus 1 
alone. This sublinearity due to rectification would also occur in a
linear threshold network and is a separate effect from effects due to
the sublinear regime of the SSN.  More generally, the results of
adding stimuli with different values of $\frac{g_E}{g_I}$ are beyond
the scope of what we consider in this paper.

Here we introduce the following  five different criteria for the
transition to the sublinearly normalizing regime of the SSN, expressed in terms of the 2-D model, and study their inter-relationships:
\begin{enumerate}

\item  The direct definition of normalization in the high-dimensional
  ring model, as it follows from the approximate 2-D reduction:
\be\label{def1}
\frac{\partial \vr}{\partial \psi}  <0 \iff \frac{\partial \ln \vr}{\partial \ln c}  <n \iff  \alpha \frac{d \vy}{d \alpha}  < \vy
\ee
where  the equivalences (demonstrated in  Section~\ref{normsec})  hold component by component. Note that each component in \req{def1}  expresses a separate condition, \ie\ the normalizing property for the $E$ and the $I$ rates, respectively. 
We can obtain an expression for $\frac{d \vy}{d \alpha}$ by looking at the variation of the fixed point equation \req{yss}, which yields 
$ (-\one  + \bPhiy \hat \J) \delta \vy + ( \hat \J \vy + \vg )^{.n}  \delta \alpha = 0$. Here, $\bPhiy$ is the diagonal matrix $n\alpha^{\frac{1}{n}} \mathrm{diag}(\vy^{1-\frac{1}{n}})  = n\beta^{-1} \mathrm{diag}(\vy^{1-\frac{1}{n}}) $. 
Using \req{yss} again, we obtain
 \be\label{dydalph}
\alpha \frac{d\vy}{d\alpha} = - (\beta^{-1}\tJ - \one)^{-1} {\vy}
 \ee
 where we defined 
 \be\label{Kdef}
 \tJ =\mat{{K}_{EE}}{-{K}_{EI}}{{K}_{IE}}{-{K}_{II}} \equiv  n\, \mathrm{diag}(\vy^{1-\frac{1}{n}}) \hat\J
 \ee
 (note that given the positivity of $\mathrm{diag}(\vy^{1-\frac{1}{n}})$, all $K_{XY}$ are positive as defined). 
  Thus \req{def1} is equivalent (component by component) to 
\be\label{def11}
\vy > - \beta (\tJ - \beta\one) ^{-1} \vy.
 \ee
 
\item The instability of the excitatory subnetwork by itself (\ie\ with $\vr_I$ frozen at  its  fixed point value).  As we saw in Section~\ref{sec:genreqr}, this can be expressed as
 \be\label{def30}
 \calJ_{EE} >0
 \ee
 where the matrix $\calJ = \T^{-1} (\beta^{-1}\tJ - \one)$,  defined in
\req{2djacob}, is the Jacobian of the 2-D flow at the fixed point.  Given that $\T$ is positive and diagonal,  \req{def30} is equivalent to 
$
\beta^{-1} K_{EE} > 1,
$
or 
\be\label{def3}
K_{EE} = n \hat{J}_{EE}\,  y_E^{1-\frac{1}{n}} > \beta.
\ee

This criterion can also be written $n\alpha^{\frac{1}{n}}\hat{J}_{EE}\,
  y_E^{1-\frac{1}{n}} >1$ or $nk^{\frac{1}{n}}W_{EE}
  r_E^{1-\frac{1}{n}}>1$. For $\Omega_E>0$, $r_E$ becomes arbitrarily
  large and this criterion will always be met for nonzero $\hat{J}_{EE}$. For $\Omega_E<0$, the
  latter form makes clear that the 
  excitatory subnetwork becomes unstable iff it is unstable at the
  maximum value of $r_E$. 
  For $\Omega_E<0$ for $n=2$, if we assume a single curve of stable fixed
  points vs.\ contrast that peaks at the 
  maximum value of $r_E$ given in Eq.~\ref{remax}, we find that the
  condition for instability 
  at the maximum of $r_E$, and thus for the excitatory network to
  become unstable, is $J_{EE} \xx_E>1$ ($\xx_E$ given by
  \req{kappaeq}). This can be reduced to the 
  condition $g_E^2 \hat J_{EE}^2 > |\Omega_E| \left(g_I \hat J_{EE} + g_E
  \hat J_{IE}\right)$ (along with the condition ${g_I^2}{g_E^2} \nege
< \negi$ required for real $\xx_E$).  This illustrates that the
excitatory subnetwork will never become unstable
for sufficiently small $J_{EE}$. However, when other conditions are
also imposed (excitatory subnetwork never becomes unstable AND $\Det\
\J>0$, $c_{\max}>0$ (Eqs.~\ref{cmax}-\ref{cmax2}), fixed point at
$c_{\max}$ stable) the 
requirements on the $J$'s and $g$'s for all of these conditions to be
satisfied become far more complex. 

\ignore{ $g_E^2 \hat J_{EE}^2 - |\Omega_E| (2 g_I \hat J_{EE} + 
\Omega_I)>0$. }

\item Local instability of the low-contrast iteration scheme, \req{lowone}, at its fixed point (a sufficient condition for its divergence). By local instability, we mean the instability of the linearization of \req{lowone} around the fixed point solution. It is seen from \req{lowone} that the Jacobian of this linearization is  exactly $\bPhiy \hat\J = \beta^{-1}\tJ $ where $\tJ$ is the matrix  defined above. The condition for stability of a linear recurrence equation is that the modulus of all eigenvalues of the (Jacobian) matrix be smaller than 1. Thus the iteration is linearly unstable around the fixed point if and only if  at least one eigenvalue of $\tJ $ has modulus larger than $\beta$:
\be\label{def4}
|\lambda_1| >\beta \qquad \mathrm{OR} \qquad  |\lambda_2| >\beta
\ee
where $\lambda_1$ and $\lambda_2$ are the eigenvalues of $\tJ $.

\item Stability of the high-contrast iteration  scheme \req{hione} at its fixed point  (a necessary condition for its convergence). 
Similarly to the previous criterion, by this condition we mean the stability of the linearization of  \req{hione} around the fixed point. The Jacobian of the right side of \req{hione} is given by $\hat\J^{-1} \mathrm{diag}(\frac{\beta}{n} \vy^{\frac{1}{n} -1}) = \hat\J^{-1} \bPhiy^{-1} = \beta \tJ^{-1}$. Thus the linearization is stable if and only if both eigenvalues of $\beta\tJ ^{-1}$ have modulus smaller than 1. Since the eigenvalues of $\tJ^{-1}$ are the inverse of the eigenvalues of $\tJ $, this is equivalent to both eigenvalues of $\tJ$ having modulus larger than $\beta$:
\be\label{def5}
|\lambda_1| >\beta \qquad \mathrm{AND} \qquad  |\lambda_2| >\beta.
\ee
In particular,  this criterion clearly implies criterion 3. 
\end{enumerate}

Even though these four criteria do not define exactly the same transition point (\ie~the smallest  value of $\alpha$, or the corresponding value of $\beta$, for which a criterion first holds), as we will now argue, they typically occur for similar values of $\alpha$ (or $\beta $) that are $O(1)$. 
First, let us consider the parameter regimes for which  $\Omega_E \propto (-\hat\J^{-1} \g)_E >0$.\footnote{As we noted after \req{Omegadef}, for positive inputs, $g_E\geq 0$ and $g_I \geq 0$, $(-\hat\J^{-1} \g)_E >0$ implies $(-\hat\J^{-1} \g)_I >0$} As we saw in Sec.~\ref {sec-hiconscaling}, in this case,  for small $\beta$, $\vy(\beta)$ asymptotically approaches the value $-\hat\J^{-1} \g$. 
As we have normalized the magnitude of $\hat \J$ and $\g$, this
$\beta$-independent asymptotic value will be typically
$O(1)$. Therefore, from \req{Kdef}, $\tJ$ also asymptotically
approaches a $\beta$-independent limit with entries, eigenvalues, and
norm of modulus $O(1)$.  Hence, the left hand sides of the
inequalities in \req{def11} and \reqs{def3}-\myref{def5} approach an
$O(1)$ positive constant value as $\beta$ decreases (equivalently, as
$\alpha$ increases), while the right hand sides asymptotically
decrease linearly in magnitude with 
$\beta$ to zero.\footnote{The linear decrease of 
  the right hand sides in  \reqs{def3}-\myref{def5} with $\beta$ is
  obvious. The right hand side of \req{def11} asymptotically (as
  $\beta\to 0$) behaves like $- \beta \tJ^{-1} \vy$, and thus
  asymptotically also decreases in magnitude  
  linearly with $\beta$ to zero.} Thus these criteria will
always be satisfied for small enough $\beta$, and we expect that the
transition (largest $\beta$ for which the criterion holds) happens for
$\beta\sim 1$. 

The argument for parameter regimes for which $\Omega_E \propto
(-\hat\J^{-1} \g)_E <0$ is less straightforward, as in this case
$-\hat\J^{-1} \g = O(1)$ does not provide a valid asymptotic value for
$\vy$. Instead, as we saw in the discussion of supersaturation in
Sec.~\ref{sec-supersat}, $y_E$ (or $r_E$) typically reaches a maximum
and then decreases with increasing $\alpha$, vanishing at a finite
value of $\alpha$ (or $\beta$). However, as long as parameters ($\hat
\J$ and $\g$) are such that (1) the maximal value of $y_E$ (or $r_E$) is
$O(1)$ or larger, so that the left-hand sides of the inequalities are
positive and of magnitude $O(1)$ or larger; and (2) the 
contrast at which $y_E$ (or $r_E$) is
maximized is large enough and thus the corresponding $\beta\equiv
\beta^{\max}$ small enough, $\beta^{\max}\ll 1$, so that the
right-hand sides of the inequalities are of magnitude $O(\beta^{\max})$; then
there will be a finite region of $\beta$'s around $\beta^{\max}$
 for which the above criteria are satisfied. Thus, as long as
supersaturation does not begin too early, we expect that transitions
according to all the above criteria typically happen for $\beta\sim
O(1)$ also in this case.\footnote{An exception to this rule  was noted
  at the beginning of Sec.~\req {solutions}. In that example, $J_{EE}$
  is atypically small, while supersaturation starts at $\alpha \sim 1$
  and $y_E$ is pushed to zero too fast, for a value  of $\alpha$
  that is not large, such that criteria 2, \ie ~the instability of the
  excitatory subnetwork, is never realized.} 
   As we noted  in Sec.~\ref{solutions},  biologically supersaturation begins at
high contrasts, \eg\ 75\% \cite{Li_Creutzfeldt84}, relative to the
contrasts (10\%-20\%) at which the transition from supralinear to sublinear
behavior occur 
\cite{Heuer_Britten02,Ohshiro_etal11,Sengpiel_etal97,Polat_etal98}.
Thus we expect that biologically relevant parameters should yield a relatively large $c^{\max}$, and a  correspondingly small $\beta^{\max}\ll 1$. 

In addition to the above criteria, we also introduce a fifth criterion for sublinear response, which is directly based on the sublinearity of the $E$ and $I$ contrast-response curves, such as those plotted in the first rows of Figs.~\ref {parameters}--\ref{parameters2d-lowc}. However, as we will see this last criterion is not always as strongly associated with the crossover that happens for $\alpha, \beta \sim O(1)$. 

\begin{enumerate}
\setcounter{enumi}{4}
\item Sublinear growth of $r_E$ and/or $r_I$ with stimulus strength $c$:
\be\label{def2}
\frac{\partial \ln \vr}{\partial \ln c}  <1    \iff \frac{d \vy}{d \alpha}  <0.
\ee
Here, the inequality on the left is the mathematical expression of sublinear growth. As for criterion 1, we have two separate conditions here, stating the sublinear growth of the $E$ and $I$ rates, respectively. To see the equivalence with the right side in \req{def2}, note that from the definitions \myref{scaling}--\myref{dimensionless} we have $\ln \vr = \ln c - \ln\psi + \ln\vy + \mathrm{const.}$, and $\ln \alpha = (n-1)\ln c + \ln \psi + \mathrm{const}$. Thus 
$
 \frac{\partial \ln \vr}{\partial \ln c}  =  1 + (n-1) \frac{d \ln \vy}{d \ln \alpha},  
$
from which (given that $n>1$) the equivalence of the left side with $\frac{d \ln \vy}{d \ln \alpha} <0$ follows. Given the positivity  of $\vy$ and $\alpha$ the latter is equivalent to $d\vy/d\alpha <0$.
Finally, it follows from \req{dydalph} that \req{def2} is equivalent component-wise to 
\be\label{def21}
 (\tJ - \beta\one)^{-1} \vy > 0.
 \ee
The reasoning used for criteria 1-4 to argue that they should typically occur for $\beta \gtrsim 1$, cannot be used for this criteria, as the left hand side of \req {def21}  has an asymptotic value $\tJ^{-1}\vy$ for small $\beta$, and this need not have a definite sign for either component. 
However, as discussed in Sec.~ \ref{sec-supersat}, when $\Omega_E<0$,
and assuming a single curve of fixed points vs.~$c$,
$r_E$ will eventually decrease with $c$, \ie~eventually
$\frac{\partial \ln \vr}{\partial \ln c}$ becomes negative, which
means that for some lower value of $c$ (or $\alpha$)  it must have
become less than unity. Thus the transition according to this
criterion should always occur when $\Omega_E<0$. 
On the other hand, for parameter regimes for which $\Omega_E>0$, criteria 2 may never be realized (as in columns C and D of Fig.~\ref{parameters}),\footnote{This is because in those parameter regimes the zeroth order solution \req{y0Jg} for the high contrast regime yields positive values for both the $E$ and $I$ components, allowing for the subleading correction $\beta\vy_1$ to be negative  without making the total $\vy = \vy_0 + \beta\vy_1$ negative, as long as contrast is large enough (\ie~$\beta$ is small enough). Now according to \req {def2} criteria 2 is equivalent to $\partial \vy/\partial \alpha <0$, or $\partial \vy/\partial \beta >0$, while to leading order $\partial \vy/\partial \beta = \vy_1$. Hence if the allowed possibility $\vy_1<0$ is realized, criteria 2 will never be. 
}
 so this criteria is the least suitable way of characterizing this transition.
\end{enumerate}

The range of stimulus strengths, $c$, corresponding to the sublinear
regime  according to each of these criteria is demonstrated in the
bottom row of Figs.~\ref{parameters} and \ref {parameters2d-lowc}, for
the five choices of the connectivity matrix $\J$ as explained in
Section~\ref {solutions} (see the figure captions for further
explanation). Moreover, vertical lines in the plots  of   Fig.~\ref
{parameters2d-lowc}  indicate the transition points from the
supralinear (low $c$) to the sublinear (high $c$) regime. The values
at the
transitions of $\alpha = k c^{n-1} \psi \| \J\|$ ( $= k \psi \| \J\| c$ for the
value $n=2$ used in the figures)
according to these criteria are also given in the caption
of Fig.~\ref {parameters2d-lowc}.  As expected, the transitions occur
for $\alpha=O(1)$.

In summary, even though the transitions according to different
criteria happen at different numerical values of $\alpha$, 
the transitions according to criteria 1-4 typically happen for $\alpha
\sim 1$, as motivated on general grounds in Section~\ref{sec:scaling}
and discussed in more detail here (with the exception that criterion 2
may not occur for 
some parameter regimes with weak $E\Rightarrow E$ connections). That
is, all occur as part of the 
overall transition from supralinear behavior in the weak-input regime
($\alpha\ll 1$) to sublinear behavior in the strong-input regime
($\alpha\gg 1$). However, 
 the transition according to
criteria 5 may never be realized if $\Omega_E >0$, and is less
suitable as a marker of the overall transition studied here.

\section{Discussion}

We have shown in studies of a 2-D system (and found in simulation
studies of higher-dimensional systems, to be presented elsewhere) that
the supralinear network will dynamically stabilize with increasing
input strength provided the $I\Rightarrow E$ and $E\Rightarrow I$
connections mediating feedback inhibition are sufficiently strong and
the inhibitory time constant is not too slow.  This dynamic
stabilization results in a change from responses scaling supralinearly
to responses scaling sublinearly with the addition of a second input.  The system can
also yield ``supersaturation'', in which excitatory firing rates reach
a peak with increasing input strengths and then decrease (as observed
biologically \cite{Ledgeway_etal05,Li_Creutzfeldt84,Tyler_Apkarian85,Peirce07},
and as also noted theoretically by
\citeA{PersiVreeswijk:2011}),
with rates ultimately decreasing to zero for large enough input
strengths (which presumably are beyond the dynamical range of
biological inputs).  The conditions for this to occur were
characterized in the 2-D system.  The strongest sublinear behavior,
and hence behavior most likely to underly biological observations in
cerebral cortex, occurs for parameters that lead to supersaturation. 
As we will show in work to be presented elsewhere (presented as
Abstracts in
\citeA{Miller_Rubin10,Rubin_Miller10,Miller_Rubin11,Rubin_Miller11}),
this framework offers a unifying explanation for a number of
processes involving multi-input integration in sensory cortex,
including normalization and surround suppression. 

Many questions remain outstanding. As some examples: within the range
of models analyzed here, can more precise results, analogous to those
obtained here for 2-dimensional models, be obtained for
higher-dimensional models, for which we only discussed general scaling
arguments?  For any dimensionality, can useful results be obtained as
to when the network is globally stable? How will diversity of network
parameters, including in 
particular of the power $n$, alter behavior?  Presumably an even
slightly larger mean $n$ for I vs.\ E cells will enormously enhance
the range of parameters that will stabilize; experiments suggest that
I cells have significantly higher powers (\citeA{Haider_etal10}, Supp.\ Fig.\
S3d).  How will cell-to-cell 
variability of $n$ affect behavior?  How will behavior be affected by
taking into account the decreased noise level, and thus increase in
$n$ \cite{Miller_Troyer02,Hansel_vanVreeswijk02}, that occurs with increasing stimulus contrast
\cite{Finn_etal07}, \ie\ with increasing input firing rate?  How will
network behavior be modified by addition of short-term synaptic
facilitation and depression (\eg\ \citeA{Fioravante_Regehr11})?  Can
analysis be done of more biophysically realistic models, such as
networks of integrate-and-fire neurons, which have an input/output
function well approximated by a power law so long as they are firing
on input fluctuations rather than the mean input
\cite{Hansel_vanVreeswijk02}?  Note that in these models, the 
noise level, which as just mentioned controls the power $n$, can itself
be determined dynamically and differ between E and I cells
(\eg\ \citeA{vanVreeswijk_Sompolinsky98,Renart_etal10}).
What can we learn as we move beyond
the steady state to network dynamics, particularly using more
realistic models that can better capture faster dynamics and that
incorporate synaptic delays?  How will the network behave when
we incorporate multiple types of inhibitory neurons
(\eg\ \citeA{Isaacson_Scanziani11}), or of excitatory neurons, each
with their own (largely still unknown) connectivity patterns and
biophysical properties?

Despite the many open questions, we believe the basic findings are
likely to be quite robust and to underly a wide range of cerebral
cortical behavior: networks of units with supralinear input/output functions
can dynamically stabilize, resulting in a transition from supralinear
to sublinear input summation.


\end{document}